\documentclass[aps, prd, preprint, onecolumn, tightenlines, notitlepage, superscriptaddress, nofootinbib, preprintnumbers, floatfix]{revtex4-2}

\usepackage{amstext}
\usepackage{amssymb}
\usepackage{amsmath}
\usepackage{graphicx}
\usepackage{hyperref}
\usepackage{url}
\usepackage{color}
\usepackage{ulem}
\usepackage[utf8]{inputenc}
\pdfoutput=1
\usepackage[x11names]{xcolor}
\usepackage{textcomp}

\usepackage{epsfig,amsfonts,mathrsfs,amsmath,amssymb,graphicx,color,slashed,multirow}
\usepackage{amsmath,latexsym,amssymb,graphicx,slashed,hyperref,color,enumerate,url,cancel,gensymb}
\usepackage{textcomp}

\hypersetup{colorlinks,citecolor= nicegreen,linkcolor= nicegreen}
\definecolor{nicered}{rgb}{0.7,0.1,0.1}
\definecolor{nicegreen}{rgb}{0.1,0.5,0.1}

\AtBeginDocument{\hypersetup{citecolor=Green3,linkcolor=Blue1,urlcolor=Blue1}}
\begin{document}

\title{{\Large Non-unitary neutrino mixing in short and long-baseline experiments}}

\author{D. V. Forero}\email{dvanegas@udem.edu.co}
\affiliation{Universidad de Medell\'{i}n, Carrera 87 N° 30 - 65 Medell\'{i}n, Colombia}

\author{C. Giunti}
\email{carlo.giunti@to.infn.it}
\affiliation{Istituto Nazionale di Fisica Nucleare (INFN), Sezione di Torino, Via P. Giuria 1, I--10125 Torino, Italy}

\author{C. A. Ternes}\email{ternes@to.infn.it}
\affiliation{Istituto Nazionale di Fisica Nucleare (INFN), Sezione di Torino, Via P. Giuria 1, I--10125 Torino, Italy}

\author{M. T{\'o}rtola}\email{mariam@ific.uv.es}
\affiliation{Departament de Física Teòrica, Universitat de València, and Instituto de F\'{i}sica Corpuscular, CSIC-Universitat de Val\`{e}ncia, 46980 Paterna, Spain}

\begin{abstract}
Non-unitary neutrino mixing in the light neutrino sector is a direct consequence of type-I seesaw neutrino mass models. 
In these models, light neutrino mixing is described by a sub-matrix of the full lepton mixing matrix and, then, it is not unitary in general. In consequence, neutrino oscillations are characterized by additional parameters, including new sources of CP violation. 
Here we perform a combined analysis of short and long-baseline neutrino oscillation data in this extended mixing scenario.
We did not find a significant deviation from unitary mixing, and the complementary data sets have been used to constrain the non-unitarity parameters.
We have also found that the T2K and NOvA tension in the determination of the Dirac CP-phase is not alleviated in the context of non-unitary neutrino mixing.
\end{abstract}

\keywords{}
\maketitle
\newpage
\tableofcontents

\section{Introduction}
\label{sec:intro}

Current neutrino oscillation data~\cite{deSalas:2020pgw,Capozzi:2017ipn,Esteban:2020cvm} implies that neutrinos are massive particles.
The smallness of the neutrino masses arises naturally in the seesaw mechanism~\cite{Minkowski:1977sc,Yanagida:1979as,Schechter:1980gr,Schechter:1981cv,mohapatra:1981yp}. The type-I seesaw mechanism requires the existence of new heavy neutral leptons. 
In this scenario, lepton mixing has to be extended to account for the new heavy states. 
Therefore, the $3\times3$ sub-matrix of the full lepton mixing matrix, that describes the mixing among the light neutrino states, is not unitary anymore. 
Although the predictions for non-unitarity in high-scale seesaw models are negligible, larger deviations from unitarity are generally expected in
low-scale type-I seesaw models, such as the inverse and linear seesaw variants~\cite{Mohapatra:1986bd,Akhmedov:1995ip,Akhmedov:1995vm,Malinsky:2005bi,Malinsky:2009gw,Malinsky:2009df}. 
For the case of very heavy neutral leptons, with masses above the electroweak scale, precision and flavour observables can constrain the  allowed  size  of non-unitarity to  the  per-mille level~\cite{Blennow:2016jkn,Escrihuela:2016ube}. 
Likewise, direct searches for heavy neutral leptons at different experiments set strong limits on the heavy-light mixing for a wide range of masses~\cite{Atre:2009rg,Drewes:2015iva}.
Here we will focus on the complementary information on non-unitarity that can be obtained from neutrino experiments. Unfortunately, current
neutrino data can only constrain deviations from unitarity up to the percent
level~\cite{Forero:2011pc,Blennow:2016jkn,Escrihuela:2016ube}. Hopefully, upcoming neutrino experiments will improve the sensitivity to the non-unitarity of the neutrino mixing matrix in the near future~\cite{Goswami:2008mi,Ge:2016xya,Miranda:2018yym,Escrihuela:2019mot,Miranda:2020syh}.

Model independent parameterizations of the non-unitary mixing matrix can be obtained under the assumption that the new neutral particles are heavy enough to not be directly produced and, therefore, do not participate in neutrino oscillations~\cite{FernandezMartinez:2007ms,Xing:2011ur,Escrihuela:2015wra}.
A convenient parameterization of the non-unitary submatrix is obtained by multiplying the standard unitary three-neutrino mixing matrix
on the left with a triangular matrix~\cite{Escrihuela:2015wra} .
This parameterization is independent of the number of new particles~\cite{Escrihuela:2015wra}.
Note that, although seesaw mechanisms with relatively light new states~\cite{Branco:2019avf} could account for the observed short-baseline anomalies~\cite{Aguilar:2001ty,Aguilar-Arevalo:2018gpe,Aguilar-Arevalo:2020nvw,Abdurashitov:2005tb,Laveder:2007zz,Giunti:2006bj,Mention:2011rk,Giunti:2019aiy,Diaz:2019fwt,Boser:2019rta,Serebrov:2018vdw,Giunti:2021iti}, here we consider only relatively heavy (but still below the electroweak scale) new states, such that the non-unitarity of the mixing matrix is mainly constrained by neutrino oscillation data.

In recent years,  lots of efforts have been put to study the effects of a possible deviation from unitarity of three-neutrino mixing~\cite{Ohlsson:2010ca,Parke:2015goa,deGouvea:2015euy,Miranda:2016wdr,Ge:2016xya,Fernandez-Martinez:2016lgt,Pas:2016qbg,Miranda:2018yym,Martinez-Soler:2018lcy,Coutinho:2019aiy,Ellis:2020hus,Chakraborty:2020brc,Hu:2020oba}. 
In particular, it was shown that the presence of such deviations can affect the sensitivity to standard neutrino oscillation parameters in current and future neutrino experiments~\cite{Meloni:2009cg,Blennow:2016jkn,Escrihuela:2016ube,Dutta:2016czj,Li:2018jgd,Soumya:2018nkw,Miranda:2019ynh}.

In this paper we perform dedicated analyses of short and long-baseline data in presence of non-unitary neutrino mixing. 
We show that a combined analysis of the data of the short-baseline appearance experiments NOMAD and NuTeV and the long-baseline experiments MINOS/MINOS+, T2K and NOvA allows us to constrain all the non-unitarity parameters. 

We also study the effects of the new source of CP violation due to non-unitary mixing on the measurement of the standard CP-violating phase $\delta$ in T2K and NOvA~\cite{Miranda:2016wdr}.
In particular, we investigate if CP violation due to non-unitarity can ease the tension between the measurements of $\delta$ in T2K and NOvA~\cite{deSalas:2020pgw,Esteban:2020cvm,Kelly:2020fkv}.

The plan of the paper is as follows:
in Section~\ref{sec:nu_mix} we summarize the notation used in the paper and  provide the expressions of the neutrino oscillation probabilities relevant for our work. In Section~\ref{sec:sbl} we discuss non-unitary neutrino mixing in short baseline experiments. The main technical details about the long-baseline experiments considered in our analysis are discussed in Section~\ref{sec:lbl}. The results of our combined analysis of short and long-baseline data are then discussed in Sections~\ref{sec:res} and \ref{sec:cp}. Finally, in Section~\ref{sec:conc} we draw our conclusions.

\section{Non-unitary neutrino mixing}
\label{sec:nu_mix}

In type-I seesaw models, which extend the light neutrino sector with several new heavy neutral leptons, the full unitary lepton mixing matrix for 3 light neutrino states and $n-3$ heavy neutral leptons is
\begin{equation}
 U^{n\times n} = 
 \begin{pmatrix}
  N & S
  \\
  V & T
 \end{pmatrix}\,.
\label{eq:Unxn}
\end{equation}
The $3\times(n-3)$ matrix $S$ and
the $(n-3)\times3$ matrix $V$ describe the mixing between light and heavy states.
The $(n-3)\times(n-3)$ matrix $T$
contains the mixing among the heavy states, while the mixing among the light neutrino states is given by the $3\times 3$ matrix
$N$, that can be written as~\cite{Escrihuela:2015wra}
\begin{equation}
 N = N^{NP}U=
 \begin{pmatrix}
  \alpha_{11} & 0 & 0
  \\
  \alpha_{21} & \alpha_{22} & 0
  \\
  \alpha_{31} & \alpha_{32} & \alpha_{33}
 \end{pmatrix}U\,.
\label{eq:N3x3}
\end{equation}
Here, $U$ is the standard unitary three-neutrino mixing matrix. 
Therefore, all the non-unitary new physics effects are encoded in the triangular matrix $N^{NP}$, which depends on  three real positive diagonal parameters $\alpha_{ii}$, and  three complex parameters $\alpha_{ij}$ ($i\neq j$), which can be decomposed in their moduli $|\alpha_{ij}|$ and their arguments, $\phi_{ij}$, which introduce new sources of CP violation.

The non-unitarity parameters can be expressed in terms of the  mixing angles of the full matrix $U^{n\times n}$. The diagonal parameters are given by
\begin{equation}
 \alpha_{ii} = c_{in}c_{in-1
 }\ldots c_{i4}\,,
\end{equation}
where $c_{ij} = \cos\theta_{ij}$ are the cosines of the new mixing angles $\theta_{ij}$ of the matrix $U^{n\times n}$
 describing the mixing between the light and heavy states. 
The non-diagonal parameters can be written as
\begin{eqnarray}
 \alpha_{21} &= c_{2n}c_{2n-1}\ldots c_{25}\eta_{24}\bar{\eta}_{14}
 +
 c_{2n}\ldots c_{26}\eta_{25}\bar{\eta}_{15}c_{14}
 +\ldots+
 \eta_{2n}\bar{\eta}_{1n}c_{1n-1}\ldots c_{14}\,,
 \\
\alpha_{32} &= c_{3n}c_{3n-1}\ldots c_{35}\eta_{34}\bar{\eta}_{24}
 +
 c_{3n}\ldots c_{36}\eta_{35}\bar{\eta}_{25}c_{24}
 +\ldots+
 \eta_{3n}\bar{\eta}_{2n}c_{2n-1}\ldots c_{24}\,,
 \\
\alpha_{31} &= c_{3n}c_{3n-1}\ldots c_{35}\eta_{34}c_{24}\bar{\eta}_{14}
 +
 c_{3n}\ldots c_{36}\eta_{35}c_{25}\bar{\eta}_{15}c_{14}
 +\ldots+
 \eta_{3n}c_{2n}\bar{\eta}_{1n}c_{1n-1}\ldots c_{14}\,,
\end{eqnarray}
with $\eta_{ij}=\sin\theta_{ij}e^{-i\delta_{ij}}$, where $\delta_{ij}$ is the CP phase associated to the angle $\theta_{ij}$ (not to be confused with $\phi_{ij}=\arg(\alpha_{ij})$, which in general depend on these $\delta$'s). The non-diagonal parameters are related to the diagonal ones through the triangular inequality~\cite{Escrihuela:2016ube}
(see Appendix~\ref{app:uneq})
\begin{equation}
 |\alpha_{ij}|
 \leq
 \sqrt{(1-\alpha_{ii}^2)(1-\alpha_{jj}^2)}\,.
 \label{eq:rel_alpha}
\end{equation}
In the following, we briefly review the oscillation probabilities relevant for the experiments discussed in this paper.
The general expression for the neutrino oscillation probability in the $\nu_\alpha$ $\to$ $\nu_\beta$ channel is given by 
\begin{eqnarray}
 P_{\alpha\beta} = 
 | ( N N^{\dagger} )_{\alpha\beta} |^2
 &-&4\sum_{k>j}\Re\left[N_{\alpha k}^*N_{\beta k}N_{\alpha j}N_{\beta j}^*\right]\sin^2\left(\frac{\Delta m_{kj}^2L}{4E}\right)\nonumber
 \\
 &+&2\sum_{k>j}\Im\left[N_{\alpha k}^*N_{\beta k}N_{\alpha j}N_{\beta j}^*\right]\sin\left(\frac{\Delta m_{kj}^2L}{2E}\right)\,.
 \label{eq:oscprob}
\end{eqnarray}
Note that the first term of the probability is not equal to $\delta_{\alpha\beta}$ as in the unitary case (it depends only on the values of the $\alpha$
parameters, as one can see from Eq.~\eqref{NNalpha1} in Appendix~\ref{app:uneq}).
This means that, in presence of non-unitary neutrino mixing, a zero-distance flavor conversion is possible.
Apart from this, the neutrino oscillation probability has the same structure as in the standard case with $U$ replaced by $N$.
In what follows, we drop terms which are cubic products of the ``small'' parameters $\sin\theta_{13}$, $\Delta m_{21}^2/\Delta m_{31}^2$ and $|\alpha_{21}|$. In this approximation, the vacuum $\nu_\mu$ disappearance probability in presence of  non-unitarity  is given by ~\cite{Escrihuela:2015wra} 
\begin{equation}
 P_{\mu\mu} = \alpha_{22}^4 P_{\mu\mu}^{\text{st}} + \alpha_{22}^3|\alpha_{21}|P_{\mu\mu}^{I_1}+2|\alpha_{21}|^2\alpha_{22}^2P_{\mu\mu}^{I_2}\,,
 \label{eq:Pmm}
\end{equation}
where $P_{\mu\mu}^{\text{st}}$ is the standard unitary oscillation probability in vacuum and the new terms are given by
\begin{eqnarray}
 P_{\mu\mu}^{I_1} = &-&8\sin\theta_{13}\sin\theta_{23}\cos2\theta_{23}\cos(\delta-\phi_{21})\sin^2\left(\frac{\Delta m_{31}^2L}{4E}\right)
 \\
 &+&
 2\cos\theta_{23}\sin2\theta_{12}\sin^2\theta_{23}\cos\phi_{21}\sin\left(\frac{\Delta m_{31}^2L}{2E}\right)\sin\left(\frac{\Delta m_{21}^2L}{2E}\right) 
\end{eqnarray}
and
\begin{equation}
 P_{\mu\mu}^{I_2}=1-2\sin^2\theta_{23}\sin^2\left(\frac{\Delta m_{31}^2L}{4E}\right)\,.
\end{equation}
The $\nu_\mu\to\nu_e$ appearance probability is given by
\begin{equation}
P_{\mu e} = 
 (\alpha_{11}\alpha_{22})^2 P^{\text{st}}_{\mu e}
+  \alpha_{11}^2 \alpha_{22}|\alpha_{21}|  P^{I}_{\mu e} 
+ \alpha_{11}^2|\alpha_{21} |^2 \,.
\label{eq:Pmue}
\end{equation}
Again, $P^{\text{st}}_{\mu e}$ is the standard unitary oscillation probability and the new term is given by
\begin{eqnarray}
P^{I}_{\mu e} & = &
-2 
   \bigg[
   \sin2\theta_{13} \sin\theta_{23} 
   \sin\left( \frac{\Delta m^2_{31}L} {4E_\nu}\right)
   \sin\left(\frac{\Delta m^2_{31}L}{4E_\nu} + \delta- \phi_{21}\right) \bigg]
\nonumber \\ 
  & + &  \cos\theta_{13} \cos\theta_{23} 
  \sin2\theta_{12}\sin\phi_{21}
   \sin\left(\frac{\Delta m^2_{21}L}{2E_\nu}\right)\,.
\end{eqnarray}
In addition to the standard parameters, the oscillation probabilities
under consideration depend on $\alpha_{22}$, $\alpha_{11}$, $|\alpha_{21}|$ and $\phi_{21}$. 
The remaining non-unitarity parameters contribute only through matter effects~\cite{Blennow:2016jkn,Escrihuela:2016ube} to the oscillation probabilities considered here.

It should be noted that, in many experiments, the spectrum at a far detector is inferred from the measured spectrum at a near detector. In this case, the oscillation probability needs to be corrected including the non-unitary effects which have already occurred at very short distances. This becomes important in the analysis of several of the experiments considered here, see Appendix~\ref{app:effective}.

Let us also remind that one can translate parameters characterizing non-unitarity in terms of other parameterizations, such as the one defining the light mixing matrix as
$ N = \left( \openone - \eta \right) U $~\cite{FernandezMartinez:2007ms,Fernandez-Martinez:2016lgt}
that is often used to study the effects of non-unitary neutrino mixing.
In this parameterization, $\eta$ is a Hermitian $3\times3$ matrix
that describes the unitarity violations.
Comparing the expressions of
$N N^\dagger$ in the two parameterizations,
one can find that, at first order of the $\eta$ parameters,
$\alpha_{11}^2 \simeq 1 - 2 \eta_{ee}$,
$\alpha_{11} \alpha_{21}^* \simeq - 2 \eta_{e\mu}$, and
$\alpha_{22}^2 + |\alpha_{21}|^2 \simeq 1 - 2 \eta_{\mu\mu}$.
Therefore,
for small unitarity violations, we have the direct approximate relations
$\alpha_{ii} \simeq 1 - \eta_{ii}$
and
$\alpha_{21}^* \simeq - 2 \eta_{e\mu}$.

\section{Non-unitary mixing at short-baseline experiments}
\label{sec:sbl}
\begin{figure}
  \centering
  \includegraphics[width=0.6\textwidth]{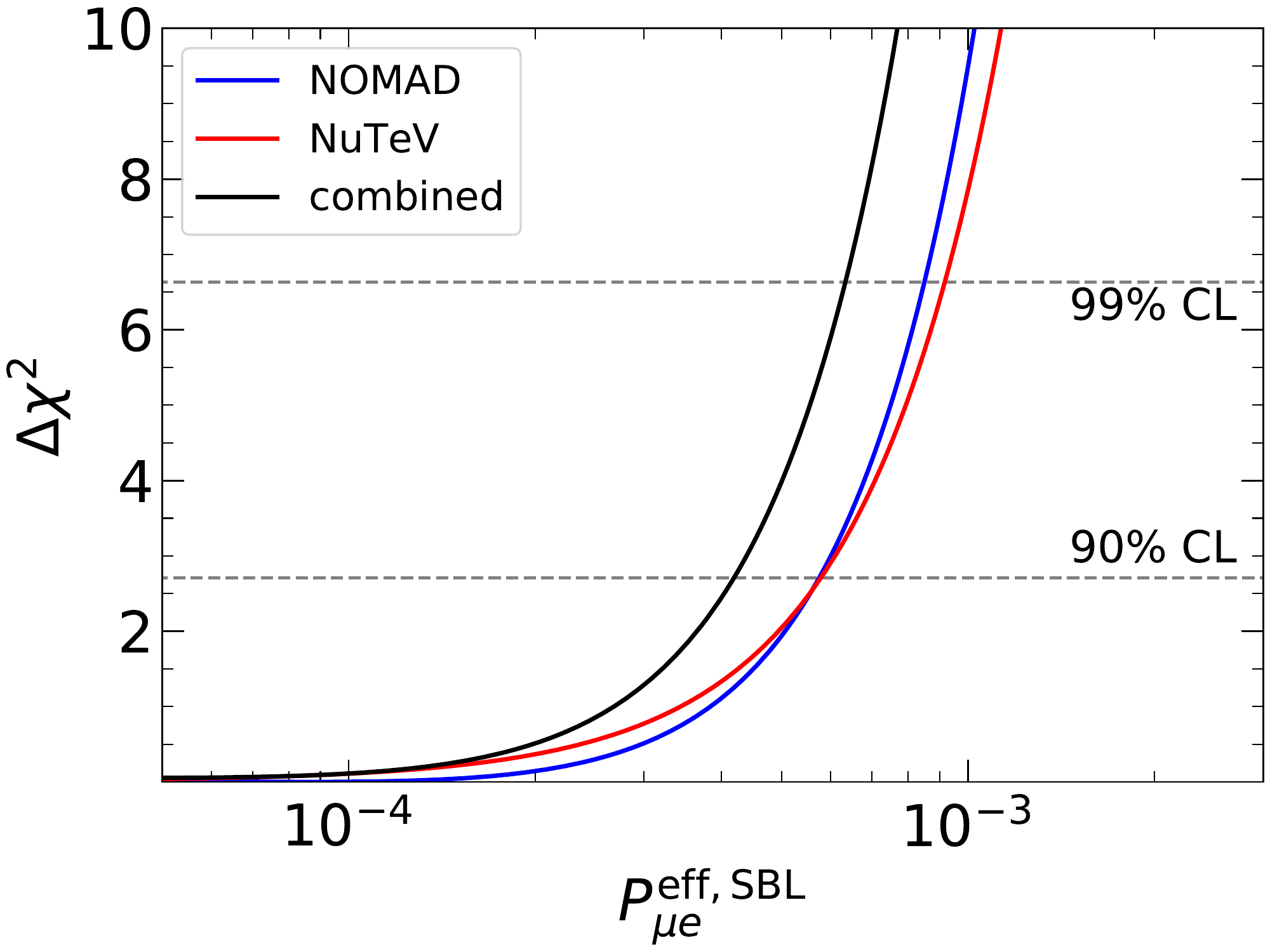}
    \caption{$\Delta\chi^2$ profiles obtained from our analysis of NOMAD (blue line) and  NuTeV (red line) data and from the combination of the two data sets (black line).}
  \label{fig:Pme_bound}
\end{figure}

In this section we discuss the effects of non-unitarity in short-baseline (SBL) $\nu_\mu\to\nu_e$ and $\bar\nu_\mu\to\bar\nu_e$ oscillation experiments and we derive the
most stringent bounds on the non-unitarity parameters that can be obtained from the current data.
We consider only these channels because other channels,
that have been considered in Ref.~\cite{Escrihuela:2016ube},
give less stringent bounds on the non-unitarity parameters
that are relevant for the combined analysis with the data of long-baseline experiments discussed in Section~\ref{sec:res}.

Considering that in the analysis of the data of short-baseline experiments
$P^{\text{st}}_{\mu e}$ and $P^{I}_{\mu e}$ in Eq.~\eqref{eq:Pmue} are negligible,
the effective probability of $\nu_\mu\to\nu_e$ and $\bar\nu_\mu\to\bar\nu_e$ transitions
takes the very simple form
\begin{equation}
 P_{\mu e}^{\text{SBL}} = \alpha_{11}^2|\alpha_{21}|^2\,.
\label{eq_Pmue_sbl} 
\end{equation}
Therefore, short-baseline experiments are only sensitive to the energy-independent zero-distance effect coming from the first term in Eq.~\eqref{eq:oscprob}.

There are several short-baseline
$\nu_\mu\to\nu_e$ and $\bar\nu_\mu\to\bar\nu_e$ oscillation experiments
that did not find any indication in favor of these transitions.
The data were analyzed using the standard unitary two-neutrino mixing approximation,
where the transition probability depends on the mixing parameter
$\sin^2 2\vartheta$
and the squared-mass difference $\Delta m^2$.
In this case,
for large values of $\Delta m^2$, oscillations are averaged and the oscillation probability
is simply equal to $\sin^2 2\vartheta / 2$.
Therefore, it is possible to obtain the bound on the probability
$P_{\mu e}^{\text{sbl}}$
in each of these short-baseline experiments from the value of the
$\chi^2$ as a function of $\sin^2 2\vartheta$ at a sufficiently large fixed value of
$\Delta m^2$.
Such bounds on $P_{\mu e}^{\text{sbl}}$
can be used to constrain the non-unitarity parameters through Eq.~\eqref{eq_Pmue_sbl}.

In the following we consider the short-baseline experiments
NOMAD~\cite{Astier:2003gs} and NuTeV~\cite{Avvakumov:2002jj},
that give the most stringent bounds on $P_{\mu e}^{\text{sbl}}$.

NOMAD was actually an experiment designed to search for short-baseline $\nu_\mu\to\nu_\tau$ appearance.
However, due to the good electron identification efficiency, it could also be used to look for short-baseline $\nu_e$ appearance from a $\nu_\mu$ beam through the charged current reaction $\nu_e + N\rightarrow e^- + X$. 
NOMAD collected data from 1995 to 1998, running principally in neutrino mode. The exposure corresponds to $5.1\times 10^{19}$ protons on target (POT) in neutrino mode and only $0.44\times 10^{19}$ in antineutrino mode.
They did not find any evidence of $\nu_\mu\to\nu_e$ oscillations.

Also the NuTeV collaboration performed a search for short-baseline $\nu_\mu\to\nu_e$ and $\bar\nu_\mu\to\bar{\nu}_e$ appearance. NuTeV used the 800~GeV proton beam from Tevatron and collected data in the time period of 1996-1997. 
The usage of focusing magnets allowed for separate analyses of $\nu_\mu\rightarrow\nu_e$ and $\bar{\nu}_\mu\rightarrow\bar{\nu}_e$.
No evidence of appearance was found for either oscillation channel. 
Here we use the results from the combined analysis of neutrino and antineutrino oscillation channels.

Note that, in the two experiments described above, the appearance signal is inferred from the $\nu_\mu$-disappearance spectrum. Therefore, the effective oscillation probability measured by these experiments is given by
(see Eq.~\eqref{PSBL} in Appendix~\ref{app:effective})
\begin{equation}
P_{\mu e}^{\text{eff,SBL}}
=
\dfrac{ \alpha_{11}^2 |\alpha_{21}|^2 }{ ( \alpha_{22}^2 + |\alpha_{21}|^2 )^2 }
,
\label{PSBLeff}
\end{equation}
instead of Eq.~\eqref{eq_Pmue_sbl}.

The bounds that can be obtained from the short-baseline NOMAD and NuTeV data are shown in Fig.~\ref{fig:Pme_bound}.
The blue (red) lines correspond to  NOMAD (NuTeV) data, while the black line is obtained  from the combination of both experiments. 
Note that short-baseline experiments cannot constrain any of the $\alpha_{ij}$ parameters independently, but only the combination of them which determines the observable transition probability in Eq.~\eqref{PSBLeff}.
We find that both experiments have similar sensitivities to the zero-distance appearance probability, obtaining $P_{\mu e}^{\text{eff,SBL}}< 6\times10^{-4}$ at 90\% C.L., while the combined bound is $P_{\mu e}^{\text{eff,SBL}}< 4 \times10^{-4} ~(6 \times10^{-4})$ at 90\% (99\%) C.L.
In Section~\ref{sec:res} we will combine  short-baseline and long-baseline neutrino data to improve the sensitivity on the non-unitary mixing. 
This combination can be easily done by transforming the $\chi^2$ results in terms of $P_{\mu e}$, as plotted in Fig.~\ref{fig:Pme_bound}, to a $\chi^2$ function depending on $\alpha_{11}$, $|\alpha_{21}|$ and $\alpha_{22}$ using Eq.~\eqref{PSBLeff}.

We considered also the NOMAD~\cite{NOMAD:2001xxt} bounds on short-baseline $\nu_{\mu}\to\nu_{\tau}$ and
$\nu_{e}\to\nu_{\tau}$ transitions\footnote{
For simplicity, we neglected the weaker limits obtained in the contemporary CHORUS experiment~\cite{CHORUS:2007wlo}
and in other previous experiments.
},
that allow us to constrain the non-unitarity parameters
$|\alpha_{31}|$ and $|\alpha_{32}|$.
Since the NOMAD signal prediction was obtained correcting the Monte Carlo by using
a sample of $\nu_{\mu}$ charged-current events from the data~\cite{NOMAD:2001xxt},
in analogy with Eq.~\eqref{PSBL} in Appendix~\ref{app:effective},
the effective oscillation probabilities are given by
\begin{align}
\null & \null
P_{\mu\tau}^{\text{eff,SBL}}
=
\dfrac
{ \left| \alpha_{22} \alpha_{32}^{*} + \alpha_{21} \alpha_{31}^{*} \right|^2 }
{ ( \alpha_{22}^2 + |\alpha_{21}|^2 )^2 }
\geq
\dfrac
{ \left( \alpha_{22} |\alpha_{32}| - |\alpha_{21}| |\alpha_{31}| \right)^2 }
{ ( \alpha_{22}^2 + |\alpha_{21}|^2 )^2 }
,
\label{PmtSBLeff}
\\
\null & \null
P_{e\tau}^{\text{eff,SBL}}
=
\dfrac{ \alpha_{11}^2 |\alpha_{31}|^2 }{ ( \alpha_{22}^2 + |\alpha_{21}|^2 )^2 }
.
\label{PetSBLeff}
\end{align}
Unfortunately,
the very complicated analysis of the NOMAD data presented in Ref.~\cite{NOMAD:2001xxt}
cannot be reproduced outside of the NOMAD collaboration.
Therefore,
we considered an approximate $\chi^2$ obtained with a linear interpolation
of the bounds published in Ref.~\cite{NOMAD:2001xxt}.

\section{Long-baseline experiments: MINOS/MINOS+, T2K and NOvA}
\label{sec:lbl}

As we showed in Section~\ref{sec:nu_mix}, if the light neutrino mixing matrix is not unitary, new correlations arise among the standard oscillation parameters and the parameters characterizing non-unitarity.
We use the MINOS/MINOS+ data sample from Ref.~\cite{Adamson:2017uda} as well as the most recent data from the long-baseline (LBL) experiments T2K~\cite{Abe:2021gky} and NOvA~\cite{alex_himmel_2020_3959581} to search for deviations from unitarity.

The Main Injector Neutrino Oscillation Search (MINOS) is an accelerator-based neutrino oscillation experiment studying muon neutrinos produced by the NuMI beam facility at Fermilab and detected at the far (near) detector located at 735~km (1.04~km) from the source. During the MINOS data taking period, the neutrino beam peaked at an energy of 3~GeV. Later, the beam was tuned to cover larger energies, with an energy peak at 7~GeV, for the upgraded version of the experiment, MINOS+. Here we consider data corresponding to an exposure of $10.56\times10^{20}$ POT in MINOS (mostly in neutrino mode, only $3.36\times10^{20}$ POT were gathered in antineutrino mode) and $5.80\times10^{20}$ POT in MINOS+ (in neutrino mode), collected in the same detectors~\cite{Adamson:2017uda}. 

The T2K collaboration observed events induced by neutrinos and antineutrinos, corresponding to an exposure at Super-Kamiokande of 1.97$\times10^{21}$ POT in neutrino mode and 1.63$\times10^{21}$ POT in antineutrino mode. 
T2K observed 318 (137) muon (anti-muon) events and 94 (16) electron (positron) events.
In addition, 14 electron events with an associated pion were recorded.
These results allowed the T2K collaboration to exclude CP-conserving values of $\delta$ at about 2$\sigma$ confidence level~\cite{patrick_dunne_2020_3959558}.

NOvA has reached 13.6$\times10^{20}$~POT in neutrino mode~\cite{NOvA:2018gge} and 12.5$\times10^{20}$~POT in antineutrino mode,
observing 212 (105) muon (anti-muon) events and 82 (33) electron (positron) events.
Unlike T2K, the latest NOvA neutrino and antineutrino data prefer values of the CP-violating phase $\delta$ close to 0.8$\pi$, in tension with the T2K result.

The most recent T2K and NOvA data as well as the relevant technical information have been extracted from
Refs.~\cite{patrick_dunne_2020_3959558} and~\cite{alex_himmel_2020_3959581}, respectively. 
For the energy reconstruction we assume Gaussian smearing adding bin-to-bin efficiencies, which are adjusted to reproduce the best-fit spectra reported by the experimental collaborations.
Our statistical analysis includes  several sources of systematic uncertainties, related to the signal and background predictions. 
We perform the analysis of the experimental data using GLoBES~\cite{Huber:2004ka,Huber:2007ji} in combination with a package which calculates the oscillation probabilities in matter in presence of non-unitary neutrino mixing, developed for the analysis in Ref.~\cite{Escrihuela:2016ube}.
 Since the spectra at the far detectors of T2K and NOvA are inferred from the measured spectra at their near detectors, the effective appearance and disappearance oscillation probabilities relevant for these experiments need to be corrected due to zero distance effects at the near detector, see Appendix~\ref{app:effective}. 
Substituting Eq.~\eqref{eq:Pmm} into Eq.~\eqref{PLBL}, we obtain the effective disappearance probability in T2K and NOvA
\begin{equation}
P_{\mu\mu}^{\text{eff,LBL}}
=
\dfrac{
\alpha_{22}^4 P_{\mu\mu}^{\text{st}} + \alpha_{22}^3|\alpha_{21}|P_{\mu\mu}^{I_1}+2|\alpha_{21}|^2\alpha_{22}^2P_{\mu\mu}^{I_2}
}{ ( \alpha_{22}^2 + |\alpha_{21}|^2 )^2 }
.
\label{PeffLBL}
\end{equation}
Since $|\alpha_{21}|$ is small,
the leading dependence on $\alpha_{22}$ of the first term in the numerator is practically cancelled by the denominator. Therefore, T2K and NOvA can not set strong constraints on $\alpha_{22}$. Likewise, the bounds on $|\alpha_{21}|$ are weak and not competitive with that of SBL experiments discussed in Section~\ref{sec:sbl}.
However, this ensures that the measurement of the standard oscillation parameters is robust in the presence of non-unitarity.

In the case of MINOS/MINOS+~\cite{Adamson:2017uda},
we adopted the analysis procedure followed by the experimental collaboration for the search of
active-sterile neutrino oscillations in Ref.~\cite{Adamson:2017uda}.
We adapted the public MINOS/MINOS+ code to account for non-unitary neutrino oscillations,
instead of active-sterile oscillations. 
In this code, the spectra at both detectors are fitted simultaneously assuming the
MINERvA flux prediction~\cite{Aliaga:2016oaz},
that was obtained with hadronic data and, hence, is independent of neutrino mixing.
Therefore, the analysis of the MINOS/MINOS+ data sample is sensitive to the zero-distance effect and allows us to put stringent bounds on the non-unitarity parameters.

Another difference with respect to the analysis of  T2K and NOvA 
is that, in the analysis of MINOS/MINOS+ data, NC events are considered in addition to CC events.
The NC sample is sensitive to the following sum of the muon neutrino survival probability
plus the electron and tau neutrino appearance probabilities,
which deviates from unity in the case of non-unitarity (and active-sterile) mixing:
\begin{equation}
    P_{\mu}^{\text{NC}}
    =
    \sum_{\alpha=e,\mu,\tau} P_{\mu\alpha}
    \approx
    \left[
    (\alpha_{11} \alpha_{22})^2\,P^{st}_{\mu e}
    + \alpha_{22}^4\,P^{st}_{\mu \mu}
    + (\alpha_{22} \alpha_{33})^2\,P^{st}_{\mu \tau}
    \right]
    \,,
\label{eq:PmuNC}
\end{equation}
where we have considered only the dominant effects
of the diagonal non-unitarity parameters,
and
$P^{st}_{\mu \alpha}$
is the standard probability of $\nu_\mu\to\nu_\alpha$ transitions in the unitary three-neutrino mixing scenario
(with
$ \sum_{\alpha=e,\mu,\tau} P^{st}_{\mu \alpha} = 1 $).
Equation~\eqref{eq:PmuNC}
shows that the analysis of MINOS/MINOS+ NC events can constrain
all the tree diagonal $\alpha$'s,
but since $\alpha_{11}$ and $\alpha_{22}$
are better constrained by MINOS/MINOS+ CC and other data,
the NC analysis is mainly relavant for constraining $\alpha_{33}$.
Moreover, using the inequality in Eq.~\eqref{eq:rel_alpha}, one can also constrain $|\alpha_{31}|$ and $|\alpha_{32}|$. 
Therefore, the analysis of the full MINOS/MINOS+ data sample allows us to fully constrain the non-unitarity of the light neutrino mixing matrix.
 
\section{Bounds on non-unitarity parameters}
\label{sec:res}

In this section we present the results of our combined analysis of short and long-baseline data in the presence of non-unitary neutrino mixing.
In the context of long-baseline neutrino oscillations, many new parameters have to be considered in the analysis. 
Regarding the standard parameters, we keep the reactor mixing angle and the solar parameters fixed at $\sin^2\theta_{13} = 0.022$, $\sin^2\theta_{12} = 0.318$ and $\Delta m_{21}^2 = 7.5\times10^{-5}$~eV$^2$, respectively~\cite{deSalas:2020pgw}.
We have checked that fixing the reactor angle or minimizing over it within its allowed 3$\sigma$-range has no effect on the results of the current analysis.
It is sufficient to consider the range determined from the unitary fit of  reactor data because the $\nu_e$ survival probability at reactor experiments is simply given by $P_{ee} = \alpha_{11}^4P_{ee}^{\text{st}}$~\cite{Escrihuela:2015wra}.
Therefore, the factor $\alpha_{11}^4$ basically takes the role of a new flux normalization and the measurement of $\theta_{13}$ using event ratios from detectors at different baselines (as done by the current reactor experiments) is robust under non-unitary deviations of neutrino mixing.
The solar parameters play only a minor role in the context of the long-baseline experiments considered here and can be safely kept fixed at their best fit values.
Regarding the non-unitarity parameters, those associated to the third row of the non-unitarity matrix $N^{NP}$, $\alpha_{3i}$,  enter the oscillation probabilities relevant for T2K and NOvA only via matter effects and their effect is very small, as shown in Ref.~\cite{Escrihuela:2016ube}. However, they can be accessed in MINOS/MINOS+ through neutral current events.
The remaining parameters ($\sin^2\theta_{23}$, $\Delta m_{31}^2$, $\delta$, $\alpha_{22}$, $\alpha_{11}$, $|\alpha_{21}|$ and $\phi_{21}$) are varied freely in the analysis of all experiments. In the case of MINOS/MINOS+, we also vary $\alpha_{33}$, that can be measured through the NC events, which are not included in the analyses of NOvA and T2K data. Using the relation in Eq.~\eqref{eq:rel_alpha} we can also bound the off-diagonal parameters, $\alpha_{3i}$\footnote{We verified that the sensitivity to these parameters comes exclusively from Eq.~\eqref{eq:rel_alpha}.}.

\begin{figure}[t!]
  \centering
  \includegraphics[width=0.45\textwidth]{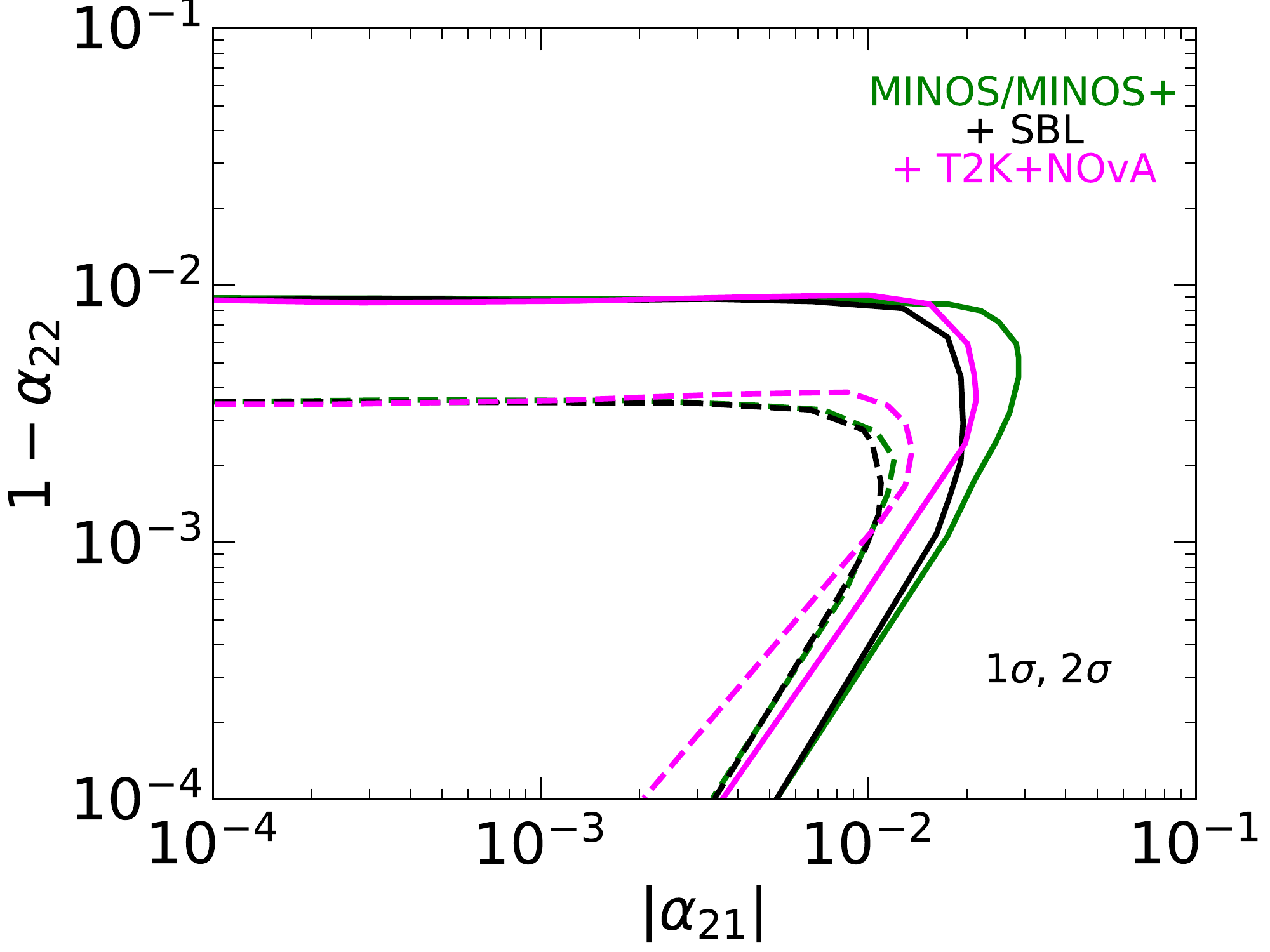}
  \includegraphics[width=0.45\textwidth]{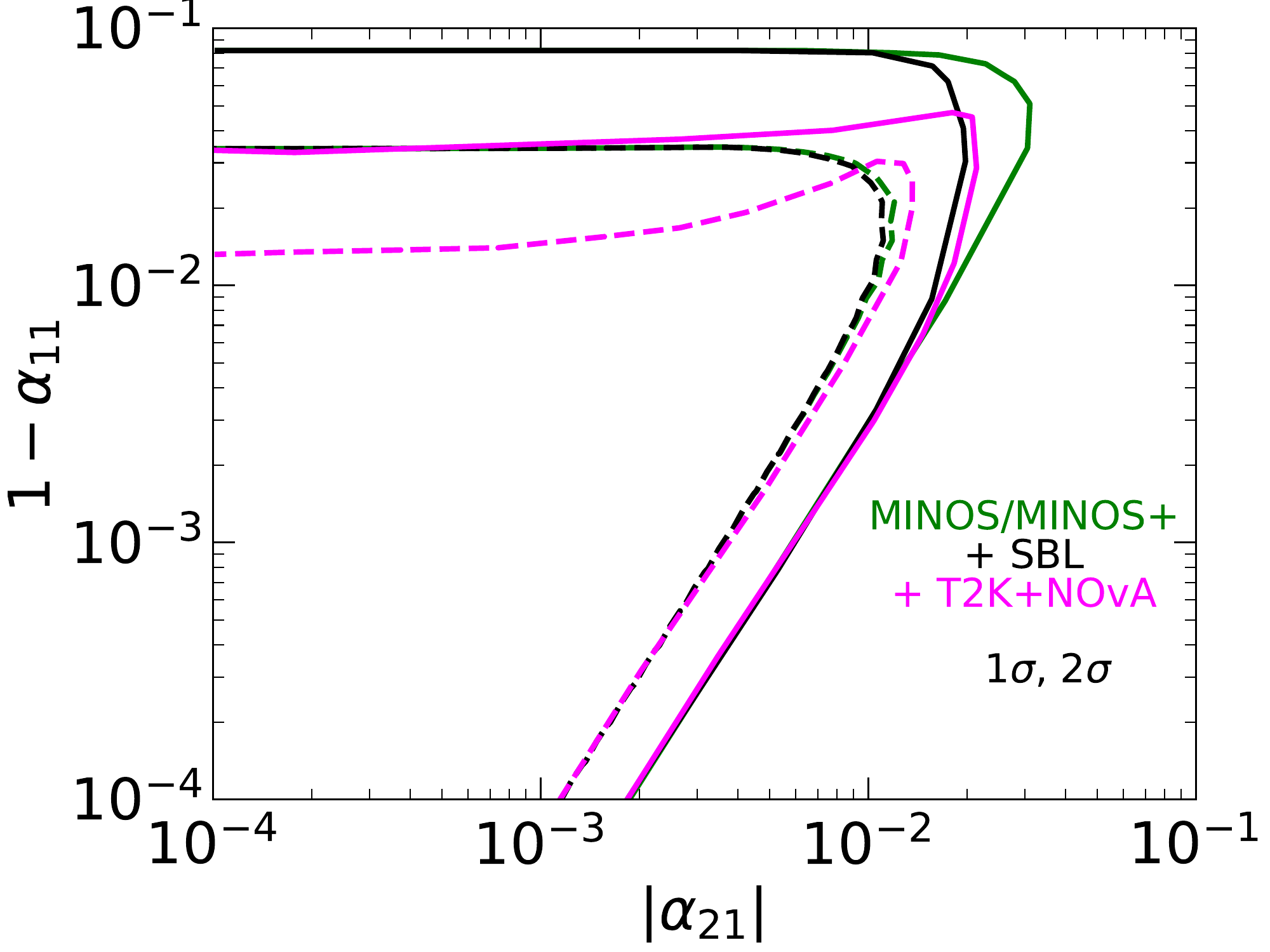}
\\  \includegraphics[width=0.45\textwidth]{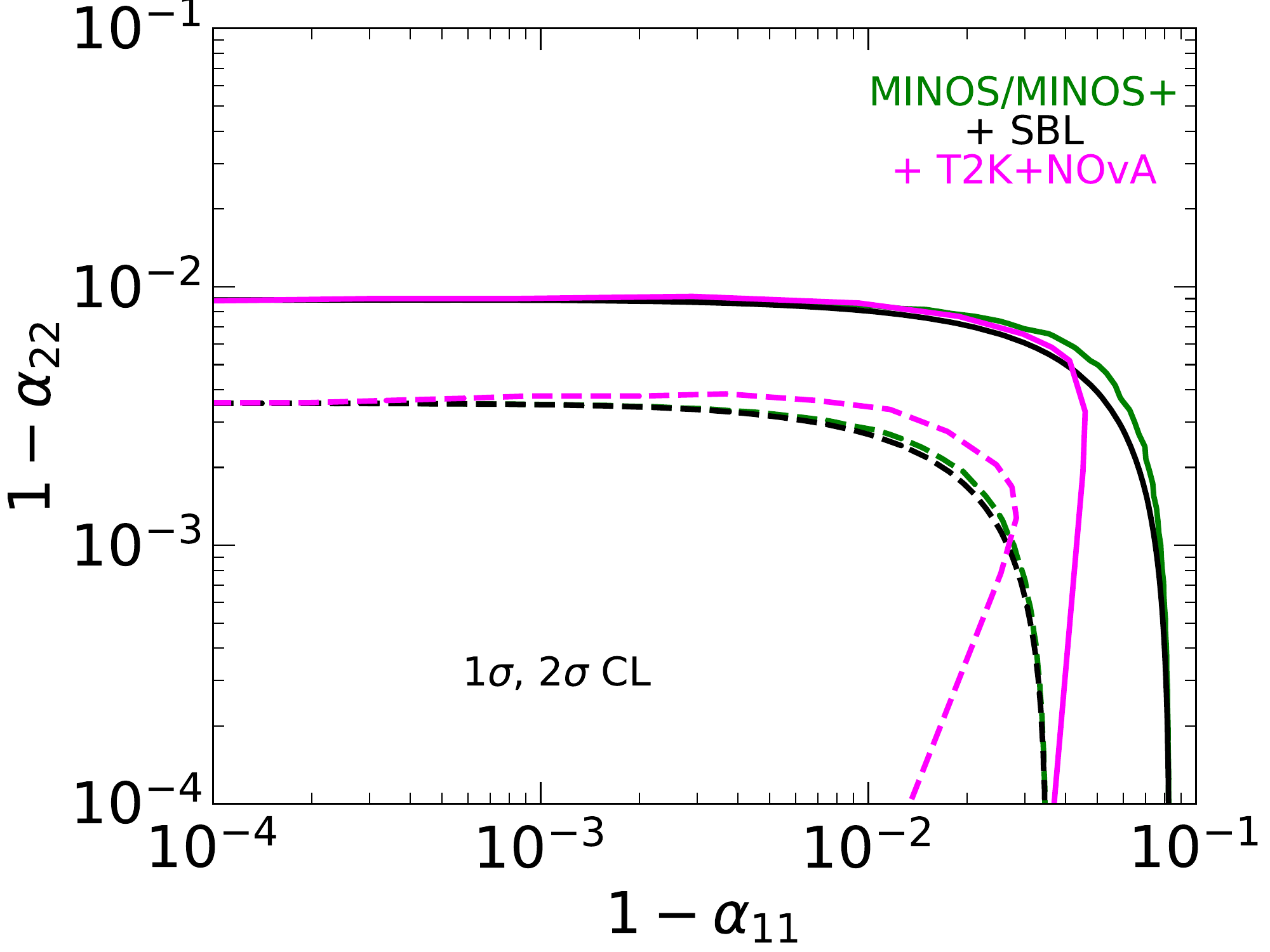}
    \caption{1$\sigma$ (dashed) and 2$\sigma$ (solid) allowed regions in   three planes of the non-unitarity parameters obtained from our analysis of MINOS/MINOS+ data (green), in combination with short-baseline oscillation data (black), and after combining with T2K and NOvA data too (magenta).}
  \label{fig:2D}
\end{figure}

Since T2K and NOvA  show a limited sensitivity to the non-unitarity parameters, we start the analysis with MINOS/MINOS+ data and then subsequently add short-baseline results and next T2K and NOvA data. 
In Fig.~\ref{fig:2D} we show the sensitivity to different non-unitarity parameters from the analysis of MINOS/MINOS+, MINOS/MINOS+ plus short-baseline experiments and the combination of all data samples. It should be noted that MINOS/MINOS+ on its own can  put strong limits on non-unitarity (see the green lines in Fig.~\ref{fig:2D}). 
Notice as well that the bound on $|\alpha_{21}|$ from MINOS/MINOS+ does not come from the
$\nu_\mu\to\nu_e$
appearance channel, which has very small statistics, but from the combined constraints on $\alpha_{11}$ and $\alpha_{22}$ and the use of the inequality in Eq.~\eqref{eq:rel_alpha}.
The addition of short-baseline data (see the black lines in Fig.~\ref{fig:2D}) has very little impact on the sensitivity to the diagonal parameters $\alpha_{11}$ and $\alpha_{22}$, but improves the bound on $|\alpha_{21}|$. 
After combining with the data from T2K and NOvA (see the magenta lines in Fig.~\ref{fig:2D}), the volumes of the allowed regions further shrink, since degeneracies among non-unitarity and standard oscillation parameters break, thanks to the better determination of the standard parameters in T2K and NOvA in comparison with MINOS/MINOS+.

\begin{figure}[t!]
  \centering
  \includegraphics[width=0.45\textwidth]{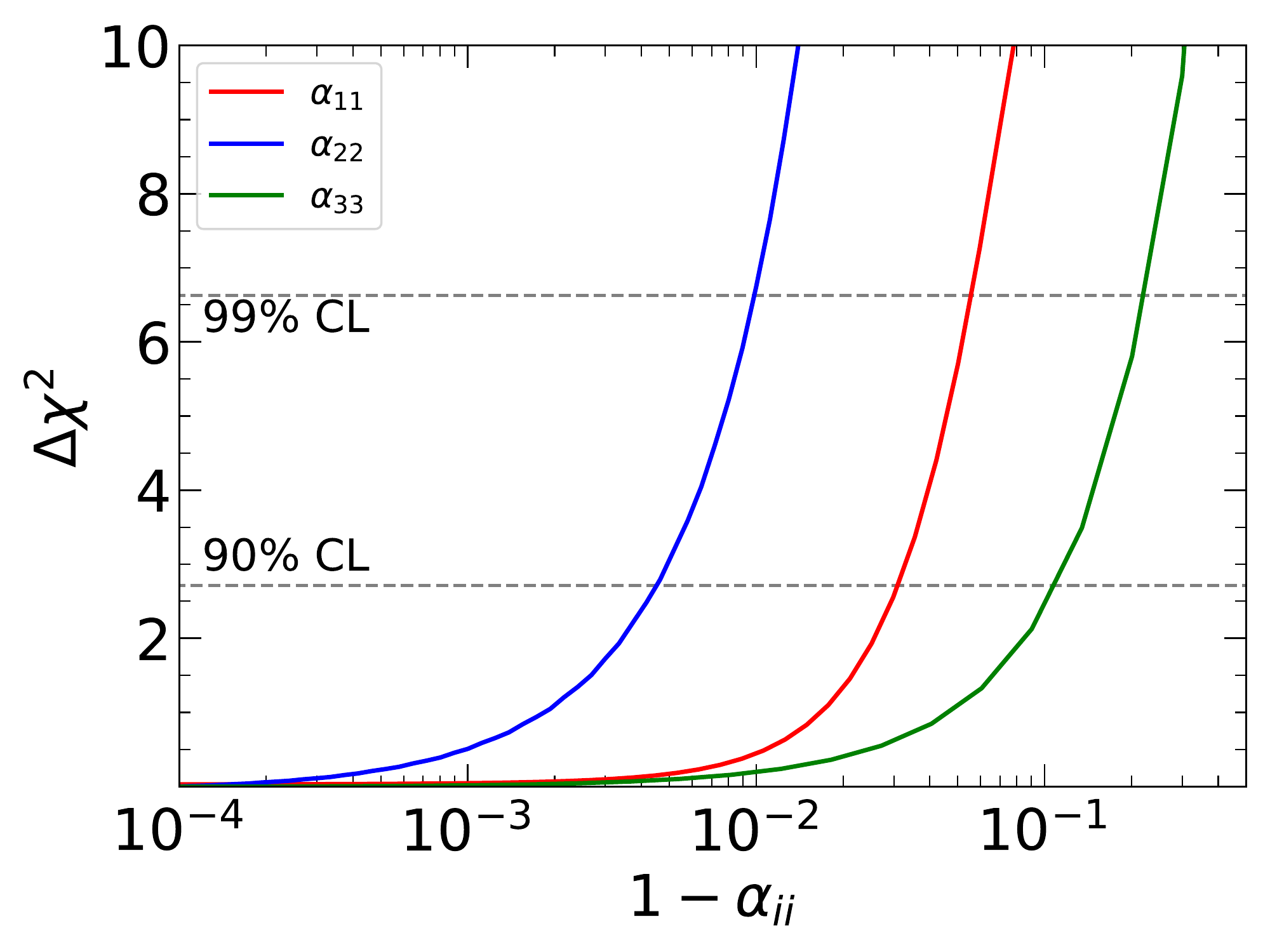}
  \includegraphics[width=0.45\textwidth]{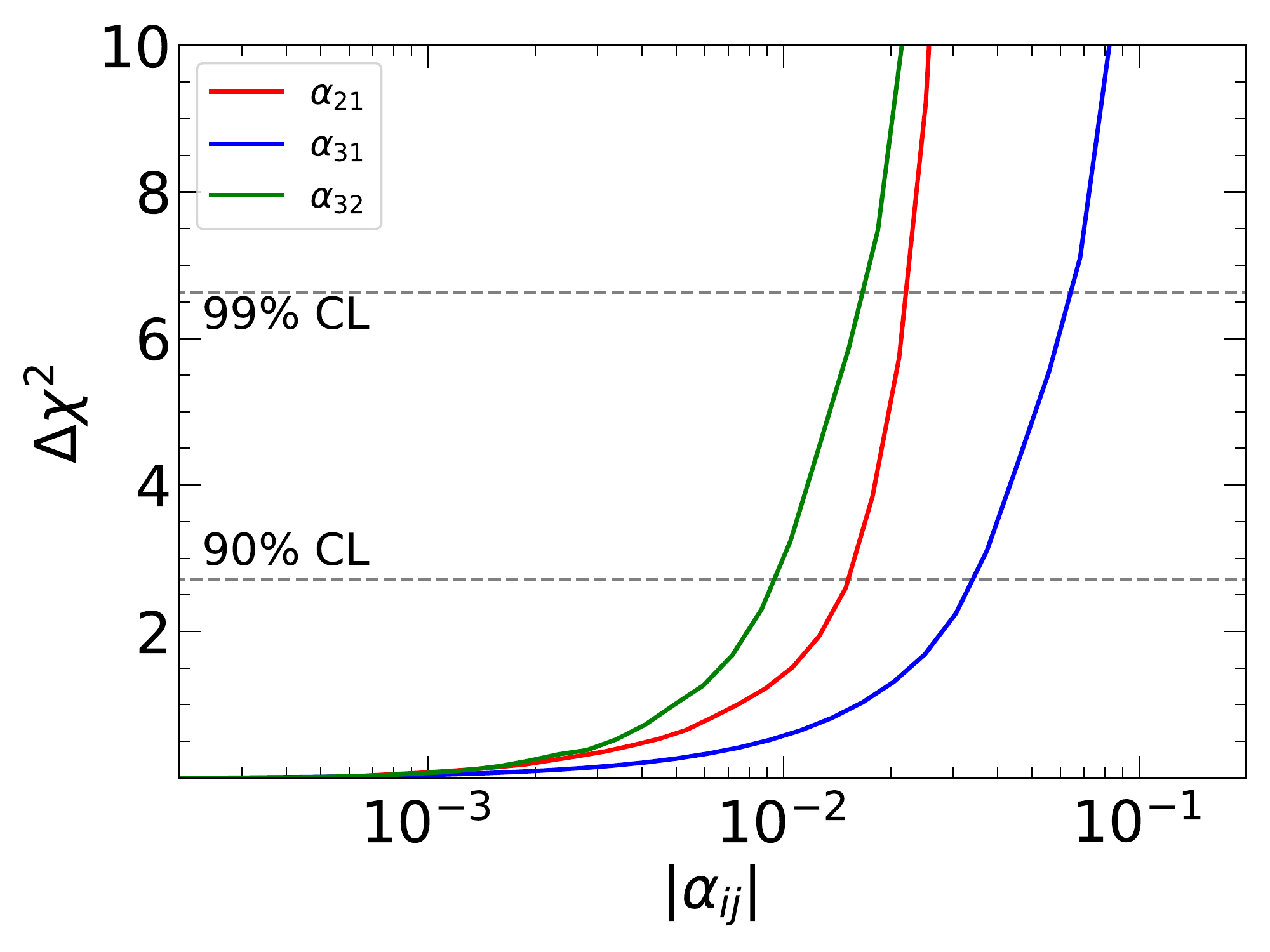}
\caption{$\Delta\chi^2$ profiles for the diagonal (left) and non-diagonal (right) non-unitarity parameters obtained from the combined analysis of short and long-baseline neutrino oscillation data.
}
  \label{fig:aXX_profiles}
\end{figure}

Figure \ref{fig:aXX_profiles} shows the
marginal $\Delta\chi^2$ profiles for
the diagonal (left) and off-diagonal (right) non-unitarity parameters obtained from the combination of all the data discussed above.
The corresponding 90\% and 99\% C.L. limits
are summarized in Tab.~\ref{tab:bounds}.
Note that some of the constraints on the non-unitarity parameters, particularly on $|\alpha_{21}|$, are slightly weaker than those in Refs.~\cite{Blennow:2016jkn,Escrihuela:2016ube,Ellis:2020hus}. This is due to the use of additional data beyond short and long-baseline results in those references and also to the consideration of the denominator in the effective short-baseline oscillation probability, Eq.~\eqref{PSBLeff}, in our current work.
However, the analysis of the MINOS/MINOS+ data assuming the MINERvA flux prediction~\cite{Aliaga:2016oaz}
allowed us to improve significantly the bound on $1-\alpha_{22}$, that is about twice as strong as in previous analyses~\cite{Escrihuela:2016ube, Blennow:2016jkn}.

As discussed at the end of Section~\ref{sec:lbl},
the analysis of MINOS/MINOS+ neutral current data
allows us to constrain $\alpha_{33}$
and,
through the inequality \eqref{eq:rel_alpha},
also $|\alpha_{31}|$ and $|\alpha_{32}|$.
Moreover, in the global analysis of short-baseline and long-baseline data, we also considered the NOMAD bounds on the non-diagonal parameters $|\alpha_{31}|$ and $|\alpha_{32}|$
discussed at the end of Section~\ref{sec:sbl}.
These results  contribute significantly to the improvement of the global bounds on the non-unitarity parameters, especially for $|\alpha_{32}|$.

\begin{table}[t!]
\centering
  \catcode`?=\active \def?{\hphantom{0}}
   \begin{tabular}{|c|l|l|}
    \hline
    Parameter & 90\% C.L. & 99\% C.L.
    \\
    \hline
    $1-\alpha_{11}$ & \, $<0.031$ & \, $<0.056$\\
    $1-\alpha_{22}$ & \, $<0.005$ & \, $<0.010$\\
    $1-\alpha_{33}$ & \, $<0.110$ & \, $<0.220$\\
    \hline
    $|\alpha_{21}|$ & \, $<0.013$ & \, $<0.023$\\
    $|\alpha_{31}|$ & \, $<0.033$ & \, $<0.065$\\
    $|\alpha_{32}|$ & \, $<0.009$ & \, $<0.017$ \\
    \hline
    \end{tabular}
    \caption{Bounds on the non-unitarity parameters obtained in this analysis.}
    \label{tab:bounds} 
\end{table}

\section{CP violation with non-unitary mixing}
\label{sec:cp}

\begin{figure}
  \centering
  \includegraphics[width=0.49\textwidth]{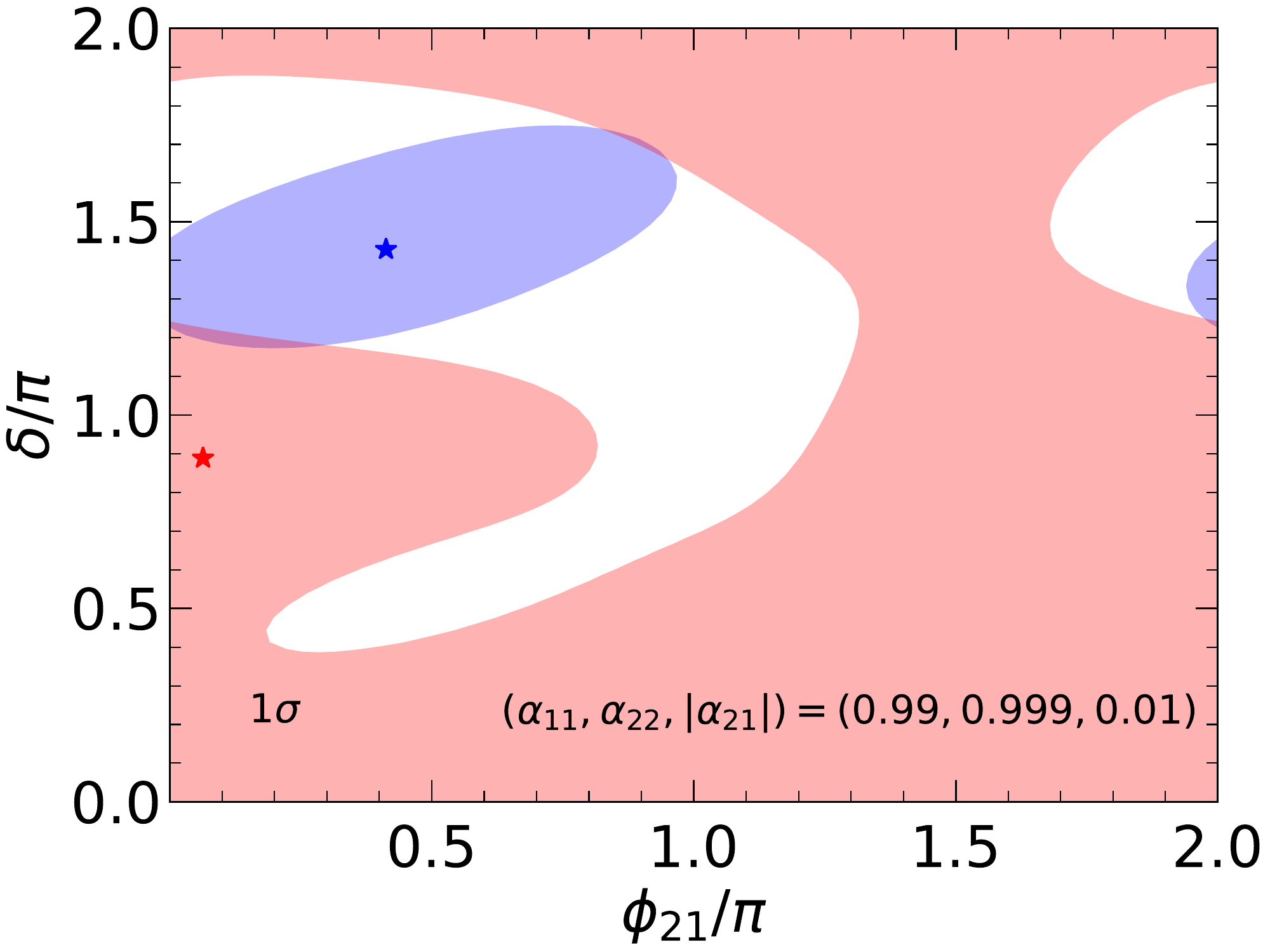}
  \includegraphics[width=0.49\textwidth]{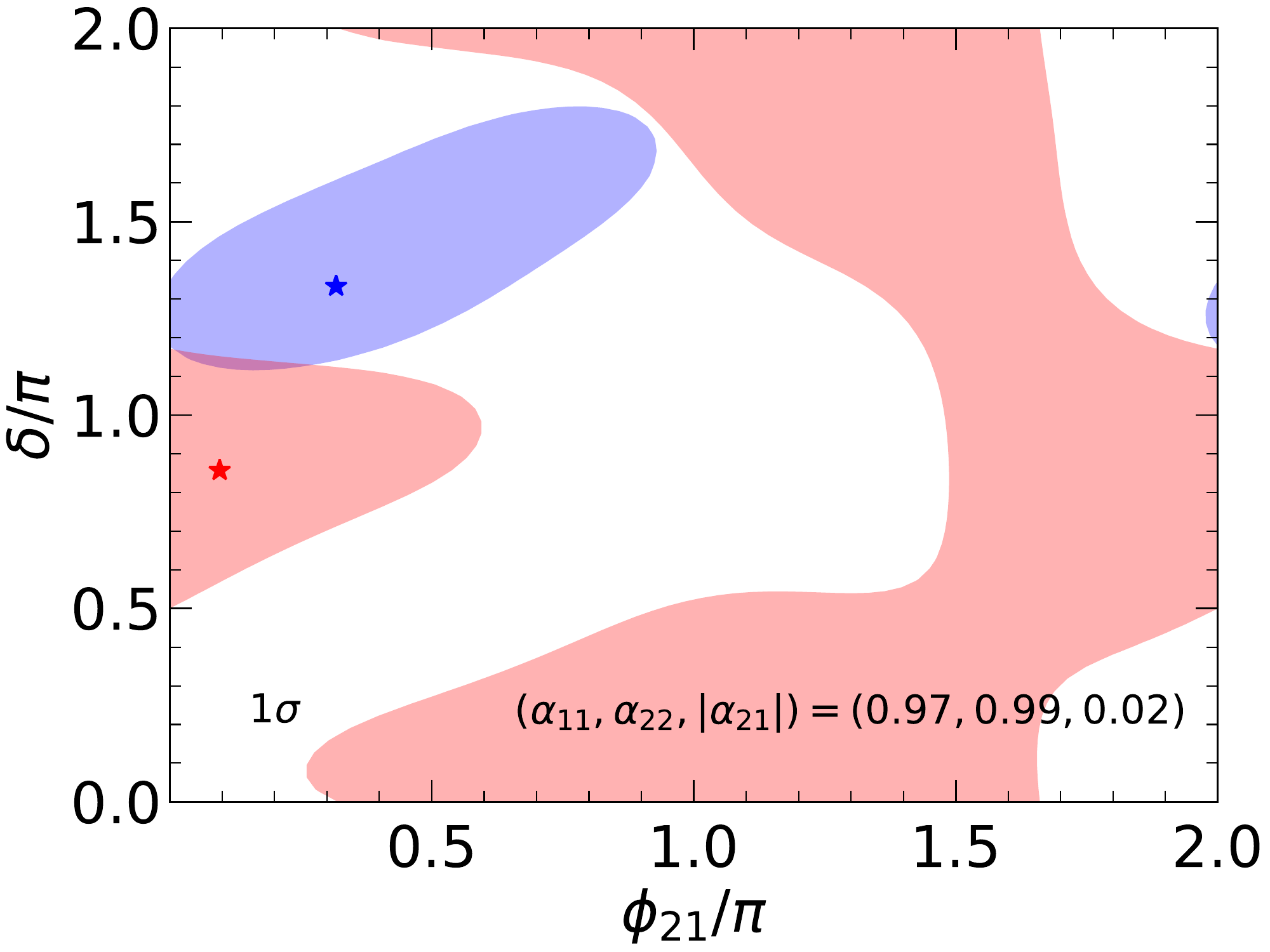}
    \caption{1$\sigma$ allowed regions in the ($\phi_{21}$, $\delta$) plane obtained from the analysis of T2K (blue regions) and NOvA (red regions). The stars indicate the corresponding best fit points. In each panel we fix $\alpha_{11}$, $\alpha_{22}$ and $|\alpha_{21}|$ to the indicated values.}
  \label{fig:del_phi21}
\end{figure}

Let us now consider the measurement of CP violation in
the T2K and NOvA long-baseline experiments
with the aim of investigating if the effects of non-unitary mixing can
resolve the tension between the data of the two experiments~\cite{deSalas:2020pgw,Esteban:2020cvm,Kelly:2020fkv}
in the case of a normal neutrino mass ordering.
The relevant CP violating phases are the standard Dirac CP phase $\delta$
and the argument  of the non-unitarity parameter $\alpha_{21}$, $\phi_{21}$.
Figure \ref{fig:del_phi21} shows the $1\sigma$
allowed regions in the $\phi_{21}$--$\delta$ plane  obtained
from the  analysis of T2K (blue regions) and NOvA (red regions)
by considering two sets of fixed values of the non-unitarity parameters
$\alpha_{11}$, $\alpha_{22}$, and $|\alpha_{21}|$.
The first set of non-unitarity parameters (left panel)
corresponds to a benchmark point well within the 90\% C.L. bounds in Tab.~\ref{tab:bounds},
while the second set (right panel) corresponds to a choice
at the borders of the 90\% and 99\% C.L. limits for $\alpha_{11}$ and $\alpha_{22}$, respectively,
and near the border of the 99\% C.L. bound for $|\alpha_{21}|$.
As one can see,
in both cases the $1\sigma$ T2K and NOvA allowed regions are almost completely disjoint,
as in the unitary case.
These two examples illustrate our general finding:
the overlap of the regions allowed by T2K and NOvA does not become significant for any combination of the $\alpha_{ij}$ parameters
that is allowed by the data.
Therefore, we conclude that, differently to what happens with other new physics scenarios~\cite{Tortola:2020ncu,Chatterjee:2020kkm,Denton:2020uda},
the non-unitarity of the neutrino mixing matrix
cannot reduce the tension between the T2K and NOvA measurements of the CP phase $\delta$
in the case of a normal neutrino mass ordering.

Let us remark that an analysis similar to that presented here has been performed in Ref.~\cite{Miranda:2019ynh}. 
The authors considered only T2K and NOvA data and obtained a preference for large deviations from unitarity, which are excluded in our analysis. In particular, we find that their best fit values, $\alpha_{11}=0.7$ and $|\alpha_{21}|=0.125$, are disfavored with very large significance, as it can be seen from Fig.~\ref{fig:aXX_profiles}. This  shows the great impact of short-baseline and MINOS/MINOS+ data in the analysis of  non-unitarity in the neutrino mixing.

\section{Conclusions}
\label{sec:conc}

The non-unitarity of the light-neutrino mixing matrix is a direct consequence of the celebrated seesaw mechanism. Therefore, analyses testing its consequences or predictions are very important for the hunt of new physics. Here we present an analysis of short and long-baseline neutrino oscillation data in the presence of non-unitary neutrino mixing.
We have found that neutrino oscillation experiments can bound some of the non-unitarity parameters at appreciable level. 
Our main results are summarized in Fig.~\ref{fig:aXX_profiles} and Tab.~\ref{tab:bounds}. 
Most of the bounds derived on the different non-unitarity parameters are comparable in size with the ones in literature~\cite{Escrihuela:2016ube,Blennow:2016jkn}. However, we can highlight a large improvement in the constraint on $1-\alpha_{22}$, which improves previous limits at least by a factor of 2. Although these results have been obtained from a combination of neutrino oscillation experiments, the largest sensitivity to non-unitarity comes from the analysis of MINOS/MINOS+ data, as shown in Fig.~\ref{fig:2D}.

We also investigated the effects of the new CP-violating phase $\phi_{21}$ on the determination of the standard CP-violating phase $\delta$ in the T2K and NOvA experiments.
In particular, we have shown that the new source of CP-violation due to non-unitarity cannot decrease the current tension between T2K and NOvA in the determination of $\delta$.

\section*{Acknowledgments}
We would like to thank Stephen Parke for useful comments on the first version of this manuscript.
CG and CAT are supported by the research grant ``The Dark Universe: A Synergic Multimessenger Approach'' number 2017X7X85K under the program ``PRIN 2017'' funded by the Ministero dell'Istruzione, Universit\`a e della Ricerca (MIUR). 
MT is supported by the Spanish grants FPA2017-85216-P (AEI/FEDER, UE),PROMETEO/2018/165 (Generalitat Valenciana) and the Spanish Red Consolider MultiDark FPA2017-90566-REDC.

\appendix

\section{Bounds on the off-diagonal $\alpha_{ij}$ parameters}
\label{app:uneq}

In this Appendix we present the proof of the
validity of the inequalities \eqref{eq:rel_alpha}
for any value of the mixing.
These inequalities were obtained in Ref.~\cite{Escrihuela:2016ube},
where they have been proved assuming small unitarity violation.

Considering the full $n \times n$ unitary matrix $U^{n\times n}$ in Eq.~\eqref{eq:Unxn},
we have the unitary relations
\begin{equation}
\sum_{k=1}^{3}
U^{n\times n}_{\alpha k}
{U^{n\times n}_{\beta k}}^{*}
+
\sum_{k=4}^{N}
U^{n\times n}_{\alpha k}
{U^{n\times n}_{\beta k}}^{*}
=
\delta_{\alpha\beta}
,
\label{uni1}
\end{equation}
that for $\alpha\neq\beta$ imply
\begin{equation}
\left|
\sum_{k=1}^{3}
U^{n\times n}_{\alpha k}
{U^{n\times n}_{\beta k}}^{*}
\right|^2
=
\left|
\sum_{k=4}^{N}
U^{n\times n}_{\alpha k}
{U^{n\times n}_{\beta k}}^{*}
\right|^2
.
\label{uni3}
\end{equation}
Applying the Cauchy-Schwarz inequality to the right-hand side of (\ref{uni3}) and using the unitarity relation (\ref{uni1}) for $\alpha=\beta$,
we obtain
\begin{align}
\left|
\sum_{k=1}^{3}
U^{n\times n}_{\alpha k}
{U^{n\times n}_{\beta k}}^{*}
\right|^2
\leq
\null & \null
\left(
\sum_{k=4}^{N}
|U^{n\times n}_{\alpha k}|^2
\right)
\left(
\sum_{k=4}^{N}
|U^{n\times n}_{\beta k}|^2
\right)
\nonumber
\\
=
\null & \null
\left(
1 - \sum_{k=1}^{3} |U^{n\times n}_{\alpha k}|^2
\right)
\left(
1 - \sum_{k=1}^{3} |U^{n\times n}_{\beta k}|^2
\right)
.
\label{uni4}
\end{align}
Considering the truncated $3\times3$ nonunitary submatrix  of $U^{n\times n}$, $N$,
the bound (\ref{uni4}) reads
\begin{equation}
| ( N N^{\dagger} )_{\alpha\beta} |^2
\leq
\left( 1 - ( N N^{\dagger} )_{\alpha\alpha} \right)
\left( 1 - ( N N^{\dagger} )_{\beta\beta} \right)
.
\label{uni5}
\end{equation}
In terms of the parameterization (\ref{eq:N3x3}) of $N$,
the matrix $N N^{\dagger}$ is given by
\begin{equation}
N N^{\dagger}
=
\begin{pmatrix}
\alpha_{11}^2
&
\alpha_{11} \alpha_{21}^{*}
&
\alpha_{11} \alpha_{31}^{*}
\\
\alpha_{11} \alpha_{21}
&
\alpha_{22}^2 + |\alpha_{21}|^2
&
\alpha_{22} \alpha_{32}^{*} + \alpha_{21} \alpha_{31}^{*}
\\
\alpha_{11} \alpha_{31}
&
\alpha_{22} \alpha_{32} + \alpha_{21}^{*} \alpha_{31}
&
\alpha_{33}^2 + |\alpha_{31}|^2 + |\alpha_{32}|^2
\end{pmatrix}
.
\label{NNalpha1}
\end{equation}
Therefore,
we have the following three inequalities:

\begin{enumerate}

\item
From
$
| ( N N^{\dagger} )_{e\mu} |^2
\leq
\left( 1 - ( N N^{\dagger} )_{ee} \right)
\left( 1 - ( N N^{\dagger} )_{\mu\mu} \right)
$
we have
\begin{equation}
\alpha_{11}^2 |\alpha_{21}|^2
\leq
\left( 1 - \alpha_{11}^2 \right)
\left( 1 - \alpha_{22}^2 - |\alpha_{21}|^2 \right)
.
\label{em1}
\end{equation}
Then,
it is straightforward to obtain the inequality (\ref{eq:rel_alpha})
for $|\alpha_{21}|$.

\item
$
| ( N N^{\dagger} )_{e\tau} |^2
\leq
\left( 1 - ( N N^{\dagger} )_{ee} \right)
\left( 1 - ( N N^{\dagger} )_{\tau\tau} \right)
$
implies that
\begin{equation}
\alpha_{11}^2 |\alpha_{31}|^2
\leq
\left( 1 - \alpha_{11}^2 \right)
\left( 1 - \alpha_{33}^2 - |\alpha_{31}|^2 - |\alpha_{32}|^2 \right)
.
\label{et1}
\end{equation}
Therefore,
\begin{equation}
|\alpha_{31}|^2
\leq
\left( 1 - \alpha_{11}^2 \right)
\left( 1 - \alpha_{33}^2 - |\alpha_{32}|^2 \right)
.
\label{et2}
\end{equation}
The obvious inequality
$
\left( 1 - \alpha_{33}^2 - |\alpha_{32}|^2 \right)
\leq
\left( 1 - \alpha_{33}^2 \right)
$
leads to the weaker constraint (\ref{eq:rel_alpha})
for $|\alpha_{31}|$.

\item
$
| ( N N^{\dagger} )_{\mu\tau} |^2
\leq
\left( 1 - ( N N^{\dagger} )_{\mu\mu} \right)
\left( 1 - ( N N^{\dagger} )_{\tau\tau} \right)
$
implies that
\begin{equation}
\left|
\alpha_{22} \alpha_{32} + \alpha_{21}^{*} \alpha_{31}
\right|^2
\leq
\left( 1 - \alpha_{22}^2 - |\alpha_{21}|^2 \right)
\left( 1 - \alpha_{33}^2 - |\alpha_{31}|^2 - |\alpha_{32}|^2 \right)
.
\label{mt1}
\end{equation}
This case is more complicated.
Since
\begin{equation}
\left|
\alpha_{22} \alpha_{32} + \alpha_{21}^{*} \alpha_{31}
\right|^2
\geq
\left(
|\alpha_{22}| |\alpha_{32}| - |\alpha_{21}| |\alpha_{31}|
\right)^2
,
\label{mt2}
\end{equation}
there are two cases that need to be considered:

\begin{enumerate}

\item
$ |\alpha_{22}| |\alpha_{32}| \leq |\alpha_{21}| |\alpha_{31}| $.
In this case there is not even need of the inequality (\ref{mt1}),
because from the inequalities
(\ref{eq:rel_alpha})
for $|\alpha_{21}|$ and $|\alpha_{31}|$
we have
\begin{align}
\alpha_{22}^2 |\alpha_{32}|^2
\leq
\null & \null
\left( 1 - \alpha_{11}^2 \right)^2
\left( 1 - \alpha_{22}^2 \right)
\left( 1 - \alpha_{33}^2 - |\alpha_{32}|^2 \right)
\nonumber
\\
\leq
\null & \null
\left( 1 - \alpha_{22}^2 \right)
\left( 1 - \alpha_{33}^2 - |\alpha_{32}|^2 \right)
,
\label{mt11}
\end{align}
that gives the inequality (\ref{eq:rel_alpha})
for $|\alpha_{32}|$.

\item
$ |\alpha_{22}| |\alpha_{32}| > |\alpha_{21}| |\alpha_{31}| $.
In this case, we have
\begin{equation}
\left|
\alpha_{22} \alpha_{32} + \alpha_{21}^{*} \alpha_{31}
\right|
\geq
|\alpha_{22}| |\alpha_{32}| - |\alpha_{21}| |\alpha_{31}|
.
\label{mt21}
\end{equation}
Therefore, from (\ref{mt1}) we obtain
\begin{equation}
|\alpha_{22}| |\alpha_{32}|
\leq
|\alpha_{21}| |\alpha_{31}|
+
\sqrt{
\left( 1 - \alpha_{22}^2 - |\alpha_{21}|^2 \right)
\left( 1 - \alpha_{33}^2 - |\alpha_{31}|^2 - |\alpha_{32}|^2 \right)
}
.
\label{mt22}
\end{equation}
The maximum of the right-hand side with respect to $|\alpha_{21}|$ and $|\alpha_{31}|$
is obtained for
\begin{equation}
|\alpha_{21}|
=
|\alpha_{31}|
\,
\sqrt{
\dfrac{ 1 - \alpha_{22}^2 }{ 1 - \alpha_{33}^2 - |\alpha_{32}|^2 }
}
.
\label{mt24}
\end{equation}
Substituting this value of $|\alpha_{21}|$ in (\ref{mt22}),
after some manipulations, we obtain
\begin{equation}
|\alpha_{22}| |\alpha_{32}|
\leq
\sqrt{
\left( 1 - \alpha_{22}^2 \right)
\left( 1 - \alpha_{33}^2 - |\alpha_{32}|^2 \right)
}
.
\label{mt25}
\end{equation}
The square of this inequality leads to the constraint (\ref{eq:rel_alpha})
for $|\alpha_{32}|$.

\end{enumerate}

\end{enumerate}

In conclusion of this Appendix,
let us remark that the inequality
(\ref{eq:rel_alpha})
for $|\alpha_{31}|$ is weaker than the constraint (\ref{et2}),
that involves also $|\alpha_{32}|$,
and
the inequality
(\ref{eq:rel_alpha})
for $|\alpha_{32}|$ is weaker than the constraint (\ref{mt1}),
that involves also $|\alpha_{21}|$, $|\alpha_{31}|$, and the relative phase between
$\alpha_{32}$ and $\alpha_{21}^{*} \alpha_{31}$.
Therefore,
in the analyses of experimental data that involve more than one of the
off-diagonal $\alpha_{ij}$ parameters
one must use the appropriate stronger constraint.

\section{Effective oscillation probabilities in SBL and LBL experiments}
\label{app:effective}

In this Appendix we derive the effective oscillation probabilities
that are measured in the
NOMAD~\cite{Astier:2003gs} and NuTeV~\cite{Avvakumov:2002jj}
short-baseline experiments
and those that are probed in our analyses of the
MINOS~\cite{Adamson:2017uda},
T2K~\cite{Abe:2021gky} and NOvA~\cite{alex_himmel_2020_3959581} long-baseline data.

Neutrino oscillation experiments that observe $\nu_{\alpha}\to\nu_{\beta}$
oscillations detect charged leptons of flavor $\beta$ that are
produced in a detector by a flux of neutrinos
produced in a source in association with charged leptons of flavor $\alpha$.
In the effective three-neutrino non-unitary mixing scheme that we are considering,
only the three light massive neutrinos are produced
in the source.
Since the effects of their sub-eV masses
can be neglected in the production and detection processes,
the number of $\nu_{\beta}$ events in a detector D
at a distance $L$ from a source of $\nu_{\alpha}$'s
is given by\footnote{
For simplicity, in this appendix we omit the geometrical $L^{-2}$
flux suppression that can be added in a straightforward way.
}
\begin{equation}
n_{\beta}^{\text{D}}
=
F^{\text{D}}_{\beta}
\,
\sigma^{\text{SM}}_{\beta}
\,
P_{\alpha\beta}(L)
\,
\Phi_{\alpha}^{\text{SM}}
,
\label{nbetaD}
\end{equation}
where the coefficient
$F^{\text{D}}_{\beta}$
takes into account all the quantities that characterize the detection processes (size, running time, efficiency, etc.),
$\sigma^{\text{SM}}_{\beta}$ is the Standard Model
charged-current weak-interaction cross section for a massless $\nu_{\beta}$,
$\Phi_{\alpha}^{\text{SM}}$ is the
flux of Standard Model massless $\nu_{\alpha}$'s produced by the source,
and
$P_{\alpha\beta}(L)$ is the oscillation probability in Eq.~(\ref{eq:oscprob}).
Then,
the flux $\Phi_{\beta}^{\text{D}}$ of $\nu_{\beta}$
obtained from the measured number of events $n_{\beta}^{\text{D}}$
considering the Standard Model
charged-current weak-interaction cross section $\sigma^{\text{SM}}_{\beta}$ is
\begin{equation}
\Phi_{\beta}^{\text{D}}
=
\dfrac{ n_{\beta}^{\text{D}} }{ F^{\text{D}}_{\beta} \, \sigma^{\text{SM}}_{\beta} }
=
P_{\alpha\beta}(L)
\,
\Phi_{\alpha}^{\text{SM}}
.
\label{PhibetaD}
\end{equation}

We analyzed the MINOS data by adapting the code in the data release of Ref.~\cite{Adamson:2017uda}
to the three-neutrino non-unitary mixing scheme.
Since the MINOS code uses the Standard Model cross sections
and the MINER$\nu$A  PPFX  flux~\cite{Aliaga:2016oaz}
obtained from hadron production data only,
the analysis corresponds to Eq.~(\ref{nbetaD})
and the effective oscillation probability coincides with
$P_{\alpha\beta}(L)$
in Eq.~(\ref{eq:oscprob}).

In the case of the T2K~\cite{Abe:2021gky} and NOvA~\cite{alex_himmel_2020_3959581},
we analyzed the far-detector (FD) $\nu_{e}$ and $\nu_{\mu}$ data considering the
$\nu_{\mu}$ flux $\Phi_{\mu}^{\text{ND}}$ measured at the near detector (ND),
where the $\nu_{\mu}$ survival probability is given by the zero-distance expression
$ P_{\mu\mu}^{0} = ( ( N N^{\dagger} )_{\mu\mu} )^2 $.
Hence, from Eq.~(\ref{PhibetaD}) the
$\nu_{\beta}$ flux $\Phi_{\beta}^{\text{FD}}$
at the far detector at the distance $L$ from the neutrino source
is given by
\begin{equation}
\Phi_{\beta}^{\text{FD}}
=
\dfrac{ P_{\mu\beta}(L) }{ ( ( N N^{\dagger} )_{\mu\mu} )^2 }
\,
\Phi_{\mu}^{\text{ND}}
\qquad
(\beta=e,\mu)
.
\label{PhiLBL}
\end{equation}
Therefore,
the effective oscillation probabilities in our analyses of the
T2K and NOvA data are
\begin{equation}
P_{\mu\beta}^{\text{eff,LBL}}
=
\dfrac{ P_{\mu\beta}(L) }{ ( ( N N^{\dagger} )_{\mu\mu} )^2 }
=
\dfrac{ P_{\mu\beta}(L) }{ ( \alpha_{22}^2 + |\alpha_{21}|^2 )^2 }
\qquad
(\beta=e,\mu)
.
\label{PLBL}
\end{equation}

Let us now consider the 
NOMAD~\cite{Astier:2003gs} and NuTeV~\cite{Avvakumov:2002jj}
SBL experiments,
that measured the ratio of $\nu_{e}$ and $\nu_{\mu}$ events
in the same detector (D) at a practically zero-distance.
In this case, the constraint on the appearance neutrino signal is inferred from the measured $\nu_\mu$ flux and, therefore, the effective
$\nu_{\mu}\to\nu_{e}$ oscillation probability
is given by the ratio of the measured $\nu_{e}$ and $\nu_{\mu}$ fluxes:
\begin{equation}
P_{\mu e}^{\text{eff,SBL}}
=
\dfrac{ \Phi_{e}^{\text{D}} }{ \Phi_{\mu}^{\text{D}} }
=
\dfrac{ P_{\mu e}^{0} }{ P_{\mu\mu}^{0} }
=
\dfrac{ | ( N N^{\dagger} )_{\mu e} |^2 }{ ( ( N N^{\dagger} )_{\mu\mu} )^2 }
=
\dfrac{ \alpha_{11}^2 |\alpha_{21}|^2 }{ ( \alpha_{22}^2 + |\alpha_{21}|^2 )^2 }
.
\label{PSBL}
\end{equation}

%

\begin{thebibliography}{72}%
\makeatletter
\providecommand \@ifxundefined [1]{%
\@ifx{#1\undefined}
}%
\providecommand \@ifnum [1]{%
\ifnum #1\expandafter \@firstoftwo
\else \expandafter \@secondoftwo
\fi
}%
\providecommand \@ifx [1]{%
\ifx #1\expandafter \@firstoftwo
\else \expandafter \@secondoftwo
\fi
}%
\providecommand \natexlab [1]{#1}%
\providecommand \enquote [1]{``#1''}%
\providecommand \bibnamefont [1]{#1}%
\providecommand \bibfnamefont [1]{#1}%
\providecommand \citenamefont [1]{#1}%
\providecommand \href@noop [0]{\@secondoftwo}%
\providecommand \href [0]{\begingroup \@sanitize@url \@href}%
\providecommand \@href[1]{\@@startlink{#1}\@@href}%
\providecommand \@@href[1]{\endgroup#1\@@endlink}%
\providecommand \@sanitize@url [0]{\catcode `\\12\catcode `\$12\catcode
`\&12\catcode `\#12\catcode `\^12\catcode `\_12\catcode `\%12\relax}%
\providecommand \@@startlink[1]{}%
\providecommand \@@endlink[0]{}%
\providecommand \url [0]{\begingroup\@sanitize@url \@url }%
\providecommand \@url [1]{\endgroup\@href {#1}{\urlprefix }}%
\providecommand \urlprefix [0]{URL }%
\providecommand \Eprint [0]{\href }%
\providecommand \doibase [0]{https://doi.org/}%
\providecommand \selectlanguage [0]{\@gobble}%
\providecommand \bibinfo [0]{\@secondoftwo}%
\providecommand \bibfield [0]{\@secondoftwo}%
\providecommand \translation [1]{[#1]}%
\providecommand \BibitemOpen [0]{}%
\providecommand \bibitemStop [0]{}%
\providecommand \bibitemNoStop [0]{.\EOS\space}%
\providecommand \EOS [0]{\spacefactor3000\relax}%
\providecommand \BibitemShut [1]{\csname bibitem#1\endcsname}%
\let\auto@bib@innerbib\@empty
\bibitem [{\citenamefont {de~Salas}\ \emph {et~al.}(2020)\citenamefont
{de~Salas}, \citenamefont {Forero}, \citenamefont {Gariazzo}, \citenamefont
{Mart\'\i{}nez-Mirav\'e}, \citenamefont {Mena}, \citenamefont {Ternes},
\citenamefont {T\'ortola},\ and\ \citenamefont {Valle}}]{deSalas:2020pgw}%
\BibitemOpen
\bibfield {author} {\bibinfo {author} {\bibfnamefont {P.~F.}\ \bibnamefont
{de~Salas}}, \bibinfo {author} {\bibfnamefont {D.~V.}\ \bibnamefont
{Forero}}, \bibinfo {author} {\bibfnamefont {S.}~\bibnamefont {Gariazzo}},
\bibinfo {author} {\bibfnamefont {P.}~\bibnamefont {Mart\'\i{}nez-Mirav\'e}},
\bibinfo {author} {\bibfnamefont {O.}~\bibnamefont {Mena}}, \bibinfo {author}
{\bibfnamefont {C.~A.}\ \bibnamefont {Ternes}}, \bibinfo {author}
{\bibfnamefont {M.}~\bibnamefont {T\'ortola}},\ and\ \bibinfo {author}
{\bibfnamefont {J.~W.~F.}\ \bibnamefont {Valle}},\ }\href
{https://doi.org/10.1007/JHEP02(2021)071} {\bibfield {journal} {\bibinfo
{journal} {JHEP}\ }\textbf {\bibinfo {volume} {21}},\ \bibinfo {pages}
{071}},\ \Eprint {https://arxiv.org/abs/2006.11237} {arXiv:2006.11237
[hep-ph]} \BibitemShut {NoStop}%
\bibitem [{\citenamefont {Capozzi}\ \emph {et~al.}(2017)\citenamefont
{Capozzi}, \citenamefont {Di~Valentino}, \citenamefont {Lisi}, \citenamefont
{Marrone}, \citenamefont {Melchiorri},\ and\ \citenamefont
{Palazzo}}]{Capozzi:2017ipn}%
\BibitemOpen
\bibfield {author} {\bibinfo {author} {\bibfnamefont {F.}~\bibnamefont
{Capozzi}}, \bibinfo {author} {\bibfnamefont {E.}~\bibnamefont
{Di~Valentino}}, \bibinfo {author} {\bibfnamefont {E.}~\bibnamefont {Lisi}},
\bibinfo {author} {\bibfnamefont {A.}~\bibnamefont {Marrone}}, \bibinfo
{author} {\bibfnamefont {A.}~\bibnamefont {Melchiorri}},\ and\ \bibinfo
{author} {\bibfnamefont {A.}~\bibnamefont {Palazzo}},\ }\href
{https://doi.org/10.1103/PhysRevD.95.096014} {\bibfield {journal} {\bibinfo
{journal} {Phys. Rev. D}\ }\textbf {\bibinfo {volume} {95}},\ \bibinfo
{pages} {096014} (\bibinfo {year} {2017})},\ \bibinfo {note} {[Addendum:
Phys.Rev.D 101, 116013 (2020)]},\ \Eprint {https://arxiv.org/abs/2003.08511}
{arXiv:2003.08511 [hep-ph]} \BibitemShut {NoStop}%
\bibitem [{\citenamefont {Esteban}\ \emph {et~al.}(2020)\citenamefont
{Esteban}, \citenamefont {Gonzalez-Garcia}, \citenamefont {Maltoni},
\citenamefont {Schwetz},\ and\ \citenamefont {Zhou}}]{Esteban:2020cvm}%
\BibitemOpen
\bibfield {author} {\bibinfo {author} {\bibfnamefont {I.}~\bibnamefont
{Esteban}}, \bibinfo {author} {\bibfnamefont {M.}~\bibnamefont
{Gonzalez-Garcia}}, \bibinfo {author} {\bibfnamefont {M.}~\bibnamefont
{Maltoni}}, \bibinfo {author} {\bibfnamefont {T.}~\bibnamefont {Schwetz}},\
and\ \bibinfo {author} {\bibfnamefont {A.}~\bibnamefont {Zhou}},\ }\href
{https://doi.org/10.1007/JHEP09(2020)178} {\bibfield {journal} {\bibinfo
{journal} {JHEP}\ }\textbf {\bibinfo {volume} {09}},\ \bibinfo {pages}
{178}},\ \Eprint {https://arxiv.org/abs/2007.14792} {arXiv:2007.14792
[hep-ph]} \BibitemShut {NoStop}%
\bibitem [{\citenamefont {Minkowski}(1977)}]{Minkowski:1977sc}%
\BibitemOpen
\bibfield {author} {\bibinfo {author} {\bibfnamefont {P.}~\bibnamefont
{Minkowski}},\ }\href {https://doi.org/10.1016/0370-2693(77)90435-X}
{\bibfield {journal} {\bibinfo {journal} {Phys. Lett. B}\ }\textbf
{\bibinfo {volume} {67}},\ \bibinfo {pages} {421} (\bibinfo {year}
{1977})}\BibitemShut {NoStop}%
\bibitem [{\citenamefont {Yanagida}(1979)}]{Yanagida:1979as}%
\BibitemOpen
\bibfield {author} {\bibinfo {author} {\bibfnamefont {T.}~\bibnamefont
{Yanagida}},\ }\href@noop {} {\bibfield {journal} {\bibinfo {journal}
{Conf. Proc. C}\ }\textbf {\bibinfo {volume} {7902131}},\ \bibinfo {pages}
{95} (\bibinfo {year} {1979})}\BibitemShut {NoStop}%
\bibitem [{\citenamefont {Schechter}\ and\ \citenamefont
{Valle}(1980)}]{Schechter:1980gr}%
\BibitemOpen
\bibfield {author} {\bibinfo {author} {\bibfnamefont {J.}~\bibnamefont
{Schechter}}\ and\ \bibinfo {author} {\bibfnamefont {J.~W.~F.}\ \bibnamefont
{Valle}},\ }\href {https://doi.org/10.1103/PhysRevD.22.2227} {\bibfield
{journal} {\bibinfo {journal} {Phys. Rev. D}\ }\textbf {\bibinfo {volume}
{22}},\ \bibinfo {pages} {2227} (\bibinfo {year} {1980})}\BibitemShut
{NoStop}%
\bibitem [{\citenamefont {Schechter}\ and\ \citenamefont
{Valle}(1982)}]{Schechter:1981cv}%
\BibitemOpen
\bibfield {author} {\bibinfo {author} {\bibfnamefont {J.}~\bibnamefont
{Schechter}}\ and\ \bibinfo {author} {\bibfnamefont {J.~W.~F.}\ \bibnamefont
{Valle}},\ }\href {https://doi.org/10.1103/PhysRevD.25.774} {\bibfield
{journal} {\bibinfo {journal} {Phys. Rev. D}\ }\textbf {\bibinfo {volume}
{25}},\ \bibinfo {pages} {774} (\bibinfo {year} {1982})}\BibitemShut
{NoStop}%
\bibitem [{\citenamefont {Mohapatra}\ and\ \citenamefont
{Senjanovic}(1981)}]{mohapatra:1981yp}%
\BibitemOpen
\bibfield {author} {\bibinfo {author} {\bibfnamefont {R.~N.}\ \bibnamefont
{Mohapatra}}\ and\ \bibinfo {author} {\bibfnamefont {G.}~\bibnamefont
{Senjanovic}},\ }\href {https://doi.org/10.1103/PhysRevD.23.165} {\bibfield
{journal} {\bibinfo {journal} {Phys. Rev. D}\ }\textbf {\bibinfo {volume}
{23}},\ \bibinfo {pages} {165} (\bibinfo {year} {1981})}\BibitemShut
{NoStop}%
\bibitem [{\citenamefont {Mohapatra}\ and\ \citenamefont
{Valle}(1986)}]{Mohapatra:1986bd}%
\BibitemOpen
\bibfield {author} {\bibinfo {author} {\bibfnamefont {R.~N.}\ \bibnamefont
{Mohapatra}}\ and\ \bibinfo {author} {\bibfnamefont {J.~W.~F.}\ \bibnamefont
{Valle}},\ }\href {https://doi.org/10.1103/PhysRevD.34.1642} {\bibfield
{journal} {\bibinfo {journal} {Phys. Rev. D}\ }\textbf {\bibinfo {volume}
{34}},\ \bibinfo {pages} {1642} (\bibinfo {year} {1986})}\BibitemShut
{NoStop}%
\bibitem [{\citenamefont {Akhmedov}\ \emph
{et~al.}(1996{\natexlab{a}})\citenamefont {Akhmedov}, \citenamefont
{Lindner}, \citenamefont {Schnapka},\ and\ \citenamefont
{Valle}}]{Akhmedov:1995ip}%
\BibitemOpen
\bibfield {author} {\bibinfo {author} {\bibfnamefont {E.~K.}\ \bibnamefont
{Akhmedov}}, \bibinfo {author} {\bibfnamefont {M.}~\bibnamefont {Lindner}},
\bibinfo {author} {\bibfnamefont {E.}~\bibnamefont {Schnapka}},\ and\
\bibinfo {author} {\bibfnamefont {J.~W.~F.}\ \bibnamefont {Valle}},\ }\href
{https://doi.org/10.1016/0370-2693(95)01504-3} {\bibfield {journal}
{\bibinfo {journal} {Phys. Lett. B}\ }\textbf {\bibinfo {volume} {368}},\
\bibinfo {pages} {270} (\bibinfo {year} {1996}{\natexlab{a}})},\ \Eprint
{https://arxiv.org/abs/hep-ph/9507275} {arXiv:hep-ph/9507275} \BibitemShut
{NoStop}%
\bibitem [{\citenamefont {Akhmedov}\ \emph
{et~al.}(1996{\natexlab{b}})\citenamefont {Akhmedov}, \citenamefont
{Lindner}, \citenamefont {Schnapka},\ and\ \citenamefont
{Valle}}]{Akhmedov:1995vm}%
\BibitemOpen
\bibfield {author} {\bibinfo {author} {\bibfnamefont {E.~K.}\ \bibnamefont
{Akhmedov}}, \bibinfo {author} {\bibfnamefont {M.}~\bibnamefont {Lindner}},
\bibinfo {author} {\bibfnamefont {E.}~\bibnamefont {Schnapka}},\ and\
\bibinfo {author} {\bibfnamefont {J.~W.~F.}\ \bibnamefont {Valle}},\ }\href
{https://doi.org/10.1103/PhysRevD.53.2752} {\bibfield {journal} {\bibinfo
{journal} {Phys. Rev. D}\ }\textbf {\bibinfo {volume} {53}},\ \bibinfo
{pages} {2752} (\bibinfo {year} {1996}{\natexlab{b}})},\ \Eprint
{https://arxiv.org/abs/hep-ph/9509255} {arXiv:hep-ph/9509255} \BibitemShut
{NoStop}%
\bibitem [{\citenamefont {Malinsky}\ \emph {et~al.}(2005)\citenamefont
{Malinsky}, \citenamefont {Romao},\ and\ \citenamefont
{Valle}}]{Malinsky:2005bi}%
\BibitemOpen
\bibfield {author} {\bibinfo {author} {\bibfnamefont {M.}~\bibnamefont
{Malinsky}}, \bibinfo {author} {\bibfnamefont {J.~C.}\ \bibnamefont
{Romao}},\ and\ \bibinfo {author} {\bibfnamefont {J.~W.~F.}\ \bibnamefont
{Valle}},\ }\href {https://doi.org/10.1103/PhysRevLett.95.161801} {\bibfield
{journal} {\bibinfo {journal} {Phys. Rev. Lett.}\ }\textbf {\bibinfo
{volume} {95}},\ \bibinfo {pages} {161801} (\bibinfo {year} {2005})},\
\Eprint {https://arxiv.org/abs/hep-ph/0506296} {arXiv:hep-ph/0506296}
\BibitemShut {NoStop}%
\bibitem [{\citenamefont {Malinsky}\ \emph
{et~al.}(2009{\natexlab{a}})\citenamefont {Malinsky}, \citenamefont
{Ohlsson},\ and\ \citenamefont {Zhang}}]{Malinsky:2009gw}%
\BibitemOpen
\bibfield {author} {\bibinfo {author} {\bibfnamefont {M.}~\bibnamefont
{Malinsky}}, \bibinfo {author} {\bibfnamefont {T.}~\bibnamefont {Ohlsson}},\
and\ \bibinfo {author} {\bibfnamefont {H.}~\bibnamefont {Zhang}},\ }\href
{https://doi.org/10.1103/PhysRevD.79.073009} {\bibfield {journal} {\bibinfo
{journal} {Phys. Rev. D}\ }\textbf {\bibinfo {volume} {79}},\ \bibinfo
{pages} {073009} (\bibinfo {year} {2009}{\natexlab{a}})},\ \Eprint
{https://arxiv.org/abs/0903.1961} {arXiv:0903.1961 [hep-ph]} \BibitemShut
{NoStop}%
\bibitem [{\citenamefont {Malinsky}\ \emph
{et~al.}(2009{\natexlab{b}})\citenamefont {Malinsky}, \citenamefont
{Ohlsson}, \citenamefont {Xing},\ and\ \citenamefont
{Zhang}}]{Malinsky:2009df}%
\BibitemOpen
\bibfield {author} {\bibinfo {author} {\bibfnamefont {M.}~\bibnamefont
{Malinsky}}, \bibinfo {author} {\bibfnamefont {T.}~\bibnamefont {Ohlsson}},
\bibinfo {author} {\bibfnamefont {Z.-z.}\ \bibnamefont {Xing}},\ and\
\bibinfo {author} {\bibfnamefont {H.}~\bibnamefont {Zhang}},\ }\href
{https://doi.org/10.1016/j.physletb.2009.07.038} {\bibfield {journal}
{\bibinfo {journal} {Phys. Lett. B}\ }\textbf {\bibinfo {volume} {679}},\
\bibinfo {pages} {242} (\bibinfo {year} {2009}{\natexlab{b}})},\ \Eprint
{https://arxiv.org/abs/0905.2889} {arXiv:0905.2889 [hep-ph]} \BibitemShut
{NoStop}%
\bibitem [{\citenamefont {Blennow}\ \emph {et~al.}(2017)\citenamefont
{Blennow}, \citenamefont {Coloma}, \citenamefont {Fernandez-Martinez},
\citenamefont {Hernandez-Garcia},\ and\ \citenamefont
{Lopez-Pavon}}]{Blennow:2016jkn}%
\BibitemOpen
\bibfield {author} {\bibinfo {author} {\bibfnamefont {M.}~\bibnamefont
{Blennow}}, \bibinfo {author} {\bibfnamefont {P.}~\bibnamefont {Coloma}},
\bibinfo {author} {\bibfnamefont {E.}~\bibnamefont {Fernandez-Martinez}},
\bibinfo {author} {\bibfnamefont {J.}~\bibnamefont {Hernandez-Garcia}},\ and\
\bibinfo {author} {\bibfnamefont {J.}~\bibnamefont {Lopez-Pavon}},\ }\href
{https://doi.org/10.1007/JHEP04(2017)153} {\bibfield {journal} {\bibinfo
{journal} {JHEP}\ }\textbf {\bibinfo {volume} {04}},\ \bibinfo {pages}
{153}},\ \Eprint {https://arxiv.org/abs/1609.08637} {arXiv:1609.08637
[hep-ph]} \BibitemShut {NoStop}%
\bibitem [{\citenamefont {Escrihuela}\ \emph {et~al.}(2017)\citenamefont
{Escrihuela}, \citenamefont {Forero}, \citenamefont {Miranda}, \citenamefont
{T\'ortola},\ and\ \citenamefont {Valle}}]{Escrihuela:2016ube}%
\BibitemOpen
\bibfield {author} {\bibinfo {author} {\bibfnamefont {F.~J.}\ \bibnamefont
{Escrihuela}}, \bibinfo {author} {\bibfnamefont {D.~V.}\ \bibnamefont
{Forero}}, \bibinfo {author} {\bibfnamefont {O.~G.}\ \bibnamefont {Miranda}},
\bibinfo {author} {\bibfnamefont {M.}~\bibnamefont {T\'ortola}},\ and\
\bibinfo {author} {\bibfnamefont {J.~W.~F.}\ \bibnamefont {Valle}},\ }\href
{https://doi.org/10.1088/1367-2630/aa79ec} {\bibfield {journal} {\bibinfo
{journal} {New J. Phys.}\ }\textbf {\bibinfo {volume} {19}},\ \bibinfo
{pages} {093005} (\bibinfo {year} {2017})},\ \Eprint
{https://arxiv.org/abs/1612.07377} {arXiv:1612.07377 [hep-ph]} \BibitemShut
{NoStop}%
\bibitem [{\citenamefont {Atre}\ \emph {et~al.}(2009)\citenamefont {Atre},
\citenamefont {Han}, \citenamefont {Pascoli},\ and\ \citenamefont
{Zhang}}]{Atre:2009rg}%
\BibitemOpen
\bibfield {author} {\bibinfo {author} {\bibfnamefont {A.}~\bibnamefont
{Atre}}, \bibinfo {author} {\bibfnamefont {T.}~\bibnamefont {Han}}, \bibinfo
{author} {\bibfnamefont {S.}~\bibnamefont {Pascoli}},\ and\ \bibinfo {author}
{\bibfnamefont {B.}~\bibnamefont {Zhang}},\ }\href
{https://doi.org/10.1088/1126-6708/2009/05/030} {\bibfield {journal}
{\bibinfo {journal} {JHEP}\ }\textbf {\bibinfo {volume} {05}},\ \bibinfo
{pages} {030}},\ \Eprint {https://arxiv.org/abs/0901.3589} {arXiv:0901.3589
[hep-ph]} \BibitemShut {NoStop}%
\bibitem [{\citenamefont {Drewes}\ and\ \citenamefont
{Garbrecht}(2017)}]{Drewes:2015iva}%
\BibitemOpen
\bibfield {author} {\bibinfo {author} {\bibfnamefont {M.}~\bibnamefont
{Drewes}}\ and\ \bibinfo {author} {\bibfnamefont {B.}~\bibnamefont
{Garbrecht}},\ }\href {https://doi.org/10.1016/j.nuclphysb.2017.05.001}
{\bibfield {journal} {\bibinfo {journal} {Nucl. Phys. B}\ }\textbf
{\bibinfo {volume} {921}},\ \bibinfo {pages} {250} (\bibinfo {year}
{2017})},\ \Eprint {https://arxiv.org/abs/1502.00477} {arXiv:1502.00477
[hep-ph]} \BibitemShut {NoStop}%
\bibitem [{\citenamefont {Forero}\ \emph {et~al.}(2011)\citenamefont {Forero},
\citenamefont {Morisi}, \citenamefont {Tortola},\ and\ \citenamefont
{Valle}}]{Forero:2011pc}%
\BibitemOpen
\bibfield {author} {\bibinfo {author} {\bibfnamefont {D.~V.}\ \bibnamefont
{Forero}}, \bibinfo {author} {\bibfnamefont {S.}~\bibnamefont {Morisi}},
\bibinfo {author} {\bibfnamefont {M.}~\bibnamefont {Tortola}},\ and\ \bibinfo
{author} {\bibfnamefont {J.~W.~F.}\ \bibnamefont {Valle}},\ }\href
{https://doi.org/10.1007/JHEP09(2011)142} {\bibfield {journal} {\bibinfo
{journal} {JHEP}\ }\textbf {\bibinfo {volume} {09}},\ \bibinfo {pages}
{142}},\ \Eprint {https://arxiv.org/abs/1107.6009} {arXiv:1107.6009 [hep-ph]}
\BibitemShut {NoStop}%
\bibitem [{\citenamefont {Goswami}\ and\ \citenamefont
{Ota}(2008)}]{Goswami:2008mi}%
\BibitemOpen
\bibfield {author} {\bibinfo {author} {\bibfnamefont {S.}~\bibnamefont
{Goswami}}\ and\ \bibinfo {author} {\bibfnamefont {T.}~\bibnamefont {Ota}},\
}\href {https://doi.org/10.1103/PhysRevD.78.033012} {\bibfield {journal}
{\bibinfo {journal} {Phys. Rev. D}\ }\textbf {\bibinfo {volume} {78}},\
\bibinfo {pages} {033012} (\bibinfo {year} {2008})},\ \Eprint
{https://arxiv.org/abs/0802.1434} {arXiv:0802.1434 [hep-ph]} \BibitemShut
{NoStop}%
\bibitem [{\citenamefont {Ge}\ \emph {et~al.}(2017)\citenamefont {Ge},
\citenamefont {Pasquini}, \citenamefont {Tortola},\ and\ \citenamefont
{Valle}}]{Ge:2016xya}%
\BibitemOpen
\bibfield {author} {\bibinfo {author} {\bibfnamefont {S.-F.}\ \bibnamefont
{Ge}}, \bibinfo {author} {\bibfnamefont {P.}~\bibnamefont {Pasquini}},
\bibinfo {author} {\bibfnamefont {M.}~\bibnamefont {Tortola}},\ and\ \bibinfo
{author} {\bibfnamefont {J.~W.~F.}\ \bibnamefont {Valle}},\ }\href
{https://doi.org/10.1103/PhysRevD.95.033005} {\bibfield {journal} {\bibinfo
{journal} {Phys. Rev. D}\ }\textbf {\bibinfo {volume} {95}},\ \bibinfo
{pages} {033005} (\bibinfo {year} {2017})},\ \Eprint
{https://arxiv.org/abs/1605.01670} {arXiv:1605.01670 [hep-ph]} \BibitemShut
{NoStop}%
\bibitem [{\citenamefont {Miranda}\ \emph {et~al.}(2018)\citenamefont
{Miranda}, \citenamefont {Pasquini}, \citenamefont {T\'ortola},\ and\
\citenamefont {Valle}}]{Miranda:2018yym}%
\BibitemOpen
\bibfield {author} {\bibinfo {author} {\bibfnamefont {O.~G.}\ \bibnamefont
{Miranda}}, \bibinfo {author} {\bibfnamefont {P.}~\bibnamefont {Pasquini}},
\bibinfo {author} {\bibfnamefont {M.}~\bibnamefont {T\'ortola}},\ and\
\bibinfo {author} {\bibfnamefont {J.~W.~F.}\ \bibnamefont {Valle}},\ }\href
{https://doi.org/10.1103/PhysRevD.97.095026} {\bibfield {journal} {\bibinfo
{journal} {Phys. Rev. D}\ }\textbf {\bibinfo {volume} {97}},\ \bibinfo
{pages} {095026} (\bibinfo {year} {2018})},\ \Eprint
{https://arxiv.org/abs/1802.02133} {arXiv:1802.02133 [hep-ph]} \BibitemShut
{NoStop}%
\bibitem [{\citenamefont {Escrihuela}\ \emph {et~al.}(2020)\citenamefont
{Escrihuela}, \citenamefont {Flores},\ and\ \citenamefont
{Miranda}}]{Escrihuela:2019mot}%
\BibitemOpen
\bibfield {author} {\bibinfo {author} {\bibfnamefont {F.~J.}\ \bibnamefont
{Escrihuela}}, \bibinfo {author} {\bibfnamefont {L.~J.}\ \bibnamefont
{Flores}},\ and\ \bibinfo {author} {\bibfnamefont {O.~G.}\ \bibnamefont
{Miranda}},\ }\href {https://doi.org/10.1016/j.physletb.2020.135241}
{\bibfield {journal} {\bibinfo {journal} {Phys. Lett. B}\ }\textbf
{\bibinfo {volume} {802}},\ \bibinfo {pages} {135241} (\bibinfo {year}
{2020})},\ \Eprint {https://arxiv.org/abs/1907.12675} {arXiv:1907.12675
[hep-ph]} \BibitemShut {NoStop}%
\bibitem [{\citenamefont {Miranda}\ \emph {et~al.}(2020)\citenamefont
{Miranda}, \citenamefont {Papoulias}, \citenamefont {Sanders}, \citenamefont
{T\'ortola},\ and\ \citenamefont {Valle}}]{Miranda:2020syh}%
\BibitemOpen
\bibfield {author} {\bibinfo {author} {\bibfnamefont {O.~G.}\ \bibnamefont
{Miranda}}, \bibinfo {author} {\bibfnamefont {D.~K.}\ \bibnamefont
{Papoulias}}, \bibinfo {author} {\bibfnamefont {O.}~\bibnamefont {Sanders}},
\bibinfo {author} {\bibfnamefont {M.}~\bibnamefont {T\'ortola}},\ and\
\bibinfo {author} {\bibfnamefont {J.~W.~F.}\ \bibnamefont {Valle}},\ }\href
{https://doi.org/10.1103/PhysRevD.102.113014} {\bibfield {journal} {\bibinfo
{journal} {Phys. Rev. D}\ }\textbf {\bibinfo {volume} {102}},\ \bibinfo
{pages} {113014} (\bibinfo {year} {2020})},\ \Eprint
{https://arxiv.org/abs/2008.02759} {arXiv:2008.02759 [hep-ph]} \BibitemShut
{NoStop}%
\bibitem [{\citenamefont {Fernandez-Martinez}\ \emph
{et~al.}(2007)\citenamefont {Fernandez-Martinez}, \citenamefont {Gavela},
\citenamefont {Lopez-Pavon},\ and\ \citenamefont
{Yasuda}}]{FernandezMartinez:2007ms}%
\BibitemOpen
\bibfield {author} {\bibinfo {author} {\bibfnamefont {E.}~\bibnamefont
{Fernandez-Martinez}}, \bibinfo {author} {\bibfnamefont {M.~B.}\ \bibnamefont
{Gavela}}, \bibinfo {author} {\bibfnamefont {J.}~\bibnamefont
{Lopez-Pavon}},\ and\ \bibinfo {author} {\bibfnamefont {O.}~\bibnamefont
{Yasuda}},\ }\href {https://doi.org/10.1016/j.physletb.2007.03.069}
{\bibfield {journal} {\bibinfo {journal} {Phys. Lett. B}\ }\textbf
{\bibinfo {volume} {649}},\ \bibinfo {pages} {427} (\bibinfo {year}
{2007})},\ \Eprint {https://arxiv.org/abs/hep-ph/0703098}
{arXiv:hep-ph/0703098} \BibitemShut {NoStop}%
\bibitem [{\citenamefont {Xing}(2012)}]{Xing:2011ur}%
\BibitemOpen
\bibfield {author} {\bibinfo {author} {\bibfnamefont {Z.-z.}\ \bibnamefont
{Xing}},\ }\href {https://doi.org/10.1103/PhysRevD.85.013008} {\bibfield
{journal} {\bibinfo {journal} {Phys. Rev. D}\ }\textbf {\bibinfo {volume}
{85}},\ \bibinfo {pages} {013008} (\bibinfo {year} {2012})},\ \Eprint
{https://arxiv.org/abs/1110.0083} {arXiv:1110.0083 [hep-ph]} \BibitemShut
{NoStop}%
\bibitem [{\citenamefont {Escrihuela}\ \emph {et~al.}(2015)\citenamefont
{Escrihuela}, \citenamefont {Forero}, \citenamefont {Miranda}, \citenamefont
{Tortola},\ and\ \citenamefont {Valle}}]{Escrihuela:2015wra}%
\BibitemOpen
\bibfield {author} {\bibinfo {author} {\bibfnamefont {F.~J.}\ \bibnamefont
{Escrihuela}}, \bibinfo {author} {\bibfnamefont {D.~V.}\ \bibnamefont
{Forero}}, \bibinfo {author} {\bibfnamefont {O.~G.}\ \bibnamefont {Miranda}},
\bibinfo {author} {\bibfnamefont {M.}~\bibnamefont {Tortola}},\ and\ \bibinfo
{author} {\bibfnamefont {J.~W.~F.}\ \bibnamefont {Valle}},\ }\href
{https://doi.org/10.1103/PhysRevD.92.053009} {\bibfield {journal} {\bibinfo
{journal} {Phys. Rev. D}\ }\textbf {\bibinfo {volume} {92}},\ \bibinfo
{pages} {053009} (\bibinfo {year} {2015})},\ \bibinfo {note} {[Erratum:
Phys.Rev.D 93, 119905 (2016)]},\ \Eprint {https://arxiv.org/abs/1503.08879}
{arXiv:1503.08879 [hep-ph]} \BibitemShut {NoStop}%
\bibitem [{\citenamefont {Branco}\ \emph {et~al.}(2020)\citenamefont {Branco},
\citenamefont {Penedo}, \citenamefont {Pereira}, \citenamefont {Rebelo},\
and\ \citenamefont {Silva-Marcos}}]{Branco:2019avf}%
\BibitemOpen
\bibfield {author} {\bibinfo {author} {\bibfnamefont {G.~C.}\ \bibnamefont
{Branco}}, \bibinfo {author} {\bibfnamefont {J.~T.}\ \bibnamefont {Penedo}},
\bibinfo {author} {\bibfnamefont {P.~M.~F.}\ \bibnamefont {Pereira}},
\bibinfo {author} {\bibfnamefont {M.~N.}\ \bibnamefont {Rebelo}},\ and\
\bibinfo {author} {\bibfnamefont {J.~I.}\ \bibnamefont {Silva-Marcos}},\
}\href {https://doi.org/10.1007/JHEP07(2020)164} {\bibfield {journal}
{\bibinfo {journal} {JHEP}\ }\textbf {\bibinfo {volume} {07}},\ \bibinfo
{pages} {164}},\ \Eprint {https://arxiv.org/abs/1912.05875} {arXiv:1912.05875
[hep-ph]} \BibitemShut {NoStop}%
\bibitem [{\citenamefont {Aguilar-Arevalo}\ \emph {et~al.}(2001)\citenamefont
{Aguilar-Arevalo} \emph {et~al.}}]{Aguilar:2001ty}%
\BibitemOpen
\bibfield {author} {\bibinfo {author} {\bibfnamefont {A.}~\bibnamefont
{Aguilar-Arevalo}} \emph {et~al.} (\bibinfo {collaboration} {LSND}),\ }\href
{https://doi.org/10.1103/PhysRevD.64.112007} {\bibfield {journal} {\bibinfo
{journal} {Phys. Rev. D}\ }\textbf {\bibinfo {volume} {64}},\ \bibinfo
{pages} {112007} (\bibinfo {year} {2001})},\ \Eprint
{https://arxiv.org/abs/hep-ex/0104049} {arXiv:hep-ex/0104049} \BibitemShut
{NoStop}%
\bibitem [{\citenamefont {Aguilar-Arevalo}\ \emph {et~al.}(2018)\citenamefont
{Aguilar-Arevalo} \emph {et~al.}}]{Aguilar-Arevalo:2018gpe}%
\BibitemOpen
\bibfield {author} {\bibinfo {author} {\bibfnamefont {A.~A.}\ \bibnamefont
{Aguilar-Arevalo}} \emph {et~al.} (\bibinfo {collaboration} {MiniBooNE}),\
}\href {https://doi.org/10.1103/PhysRevLett.121.221801} {\bibfield {journal}
{\bibinfo {journal} {Phys. Rev. Lett.}\ }\textbf {\bibinfo {volume} {121}},\
\bibinfo {pages} {221801} (\bibinfo {year} {2018})},\ \Eprint
{https://arxiv.org/abs/1805.12028} {arXiv:1805.12028 [hep-ex]} \BibitemShut
{NoStop}%
\bibitem [{\citenamefont {Aguilar-Arevalo}\ \emph {et~al.}(2021)\citenamefont
{Aguilar-Arevalo} \emph {et~al.}}]{Aguilar-Arevalo:2020nvw}%
\BibitemOpen
\bibfield {author} {\bibinfo {author} {\bibfnamefont {A.~A.}\ \bibnamefont
{Aguilar-Arevalo}} \emph {et~al.} (\bibinfo {collaboration} {MiniBooNE}),\
}\href {https://doi.org/10.1103/PhysRevD.103.052002} {\bibfield {journal}
{\bibinfo {journal} {Phys. Rev. D}\ }\textbf {\bibinfo {volume} {103}},\
\bibinfo {pages} {052002} (\bibinfo {year} {2021})},\ \Eprint
{https://arxiv.org/abs/2006.16883} {arXiv:2006.16883 [hep-ex]} \BibitemShut
{NoStop}%
\bibitem [{\citenamefont {Abdurashitov}\ \emph {et~al.}(2006)\citenamefont
{Abdurashitov} \emph {et~al.}}]{Abdurashitov:2005tb}%
\BibitemOpen
\bibfield {author} {\bibinfo {author} {\bibfnamefont {J.~N.}\ \bibnamefont
{Abdurashitov}} \emph {et~al.},\ }\href
{https://doi.org/10.1103/PhysRevC.73.045805} {\bibfield {journal} {\bibinfo
{journal} {Phys. Rev. C}\ }\textbf {\bibinfo {volume} {73}},\ \bibinfo
{pages} {045805} (\bibinfo {year} {2006})},\ \Eprint
{https://arxiv.org/abs/nucl-ex/0512041} {arXiv:nucl-ex/0512041} \BibitemShut
{NoStop}%
\bibitem [{\citenamefont {Laveder}(2007)}]{Laveder:2007zz}%
\BibitemOpen
\bibfield {author} {\bibinfo {author} {\bibfnamefont {M.}~\bibnamefont
{Laveder}},\ }\href {https://doi.org/10.1016/j.nuclphysbps.2007.02.037}
{\bibfield {journal} {\bibinfo {journal} {Nucl. Phys. B Proc. Suppl.}\
}\textbf {\bibinfo {volume} {168}},\ \bibinfo {pages} {344} (\bibinfo {year}
{2007})}\BibitemShut {NoStop}%
\bibitem [{\citenamefont {Giunti}\ and\ \citenamefont
{Laveder}(2007)}]{Giunti:2006bj}%
\BibitemOpen
\bibfield {author} {\bibinfo {author} {\bibfnamefont {C.}~\bibnamefont
{Giunti}}\ and\ \bibinfo {author} {\bibfnamefont {M.}~\bibnamefont
{Laveder}},\ }\href {https://doi.org/10.1142/S0217732307025455} {\bibfield
{journal} {\bibinfo {journal} {Mod. Phys. Lett. A}\ }\textbf {\bibinfo
{volume} {22}},\ \bibinfo {pages} {2499} (\bibinfo {year} {2007})},\ \Eprint
{https://arxiv.org/abs/hep-ph/0610352} {arXiv:hep-ph/0610352} \BibitemShut
{NoStop}%
\bibitem [{\citenamefont {Mention}\ \emph {et~al.}(2011)\citenamefont
{Mention}, \citenamefont {Fechner}, \citenamefont {Lasserre}, \citenamefont
{Mueller}, \citenamefont {Lhuillier}, \citenamefont {Cribier},\ and\
\citenamefont {Letourneau}}]{Mention:2011rk}%
\BibitemOpen
\bibfield {author} {\bibinfo {author} {\bibfnamefont {G.}~\bibnamefont
{Mention}}, \bibinfo {author} {\bibfnamefont {M.}~\bibnamefont {Fechner}},
\bibinfo {author} {\bibfnamefont {T.}~\bibnamefont {Lasserre}}, \bibinfo
{author} {\bibfnamefont {T.~A.}\ \bibnamefont {Mueller}}, \bibinfo {author}
{\bibfnamefont {D.}~\bibnamefont {Lhuillier}}, \bibinfo {author}
{\bibfnamefont {M.}~\bibnamefont {Cribier}},\ and\ \bibinfo {author}
{\bibfnamefont {A.}~\bibnamefont {Letourneau}},\ }\href
{https://doi.org/10.1103/PhysRevD.83.073006} {\bibfield {journal} {\bibinfo
{journal} {Phys. Rev. D}\ }\textbf {\bibinfo {volume} {83}},\ \bibinfo
{pages} {073006} (\bibinfo {year} {2011})},\ \Eprint
{https://arxiv.org/abs/1101.2755} {arXiv:1101.2755 [hep-ex]} \BibitemShut
{NoStop}%
\bibitem [{\citenamefont {Giunti}\ and\ \citenamefont
{Lasserre}(2019)}]{Giunti:2019aiy}%
\BibitemOpen
\bibfield {author} {\bibinfo {author} {\bibfnamefont {C.}~\bibnamefont
{Giunti}}\ and\ \bibinfo {author} {\bibfnamefont {T.}~\bibnamefont
{Lasserre}},\ }\href {https://doi.org/10.1146/annurev-nucl-101918-023755}
{\bibfield {journal} {\bibinfo {journal} {Ann. Rev. Nucl. Part. Sci.}\
}\textbf {\bibinfo {volume} {69}},\ \bibinfo {pages} {163} (\bibinfo {year}
{2019})},\ \Eprint {https://arxiv.org/abs/1901.08330} {arXiv:1901.08330
[hep-ph]} \BibitemShut {NoStop}%
\bibitem [{\citenamefont {Diaz}\ \emph {et~al.}(2020)\citenamefont {Diaz},
\citenamefont {Arg\"uelles}, \citenamefont {Collin}, \citenamefont {Conrad},\
and\ \citenamefont {Shaevitz}}]{Diaz:2019fwt}%
\BibitemOpen
\bibfield {author} {\bibinfo {author} {\bibfnamefont {A.}~\bibnamefont
{Diaz}}, \bibinfo {author} {\bibfnamefont {C.~A.}\ \bibnamefont
{Arg\"uelles}}, \bibinfo {author} {\bibfnamefont {G.~H.}\ \bibnamefont
{Collin}}, \bibinfo {author} {\bibfnamefont {J.~M.}\ \bibnamefont {Conrad}},\
and\ \bibinfo {author} {\bibfnamefont {M.~H.}\ \bibnamefont {Shaevitz}},\
}\href {https://doi.org/10.1016/j.physrep.2020.08.005} {\bibfield {journal}
{\bibinfo {journal} {Phys. Rept.}\ }\textbf {\bibinfo {volume} {884}},\
\bibinfo {pages} {1} (\bibinfo {year} {2020})},\ \Eprint
{https://arxiv.org/abs/1906.00045} {arXiv:1906.00045 [hep-ex]} \BibitemShut
{NoStop}%
\bibitem [{\citenamefont {B\"oser}\ \emph {et~al.}(2020)\citenamefont
{B\"oser}, \citenamefont {Buck}, \citenamefont {Giunti}, \citenamefont
{Lesgourgues}, \citenamefont {Ludhova}, \citenamefont {Mertens},
\citenamefont {Schukraft},\ and\ \citenamefont {Wurm}}]{Boser:2019rta}%
\BibitemOpen
\bibfield {author} {\bibinfo {author} {\bibfnamefont {S.}~\bibnamefont
{B\"oser}}, \bibinfo {author} {\bibfnamefont {C.}~\bibnamefont {Buck}},
\bibinfo {author} {\bibfnamefont {C.}~\bibnamefont {Giunti}}, \bibinfo
{author} {\bibfnamefont {J.}~\bibnamefont {Lesgourgues}}, \bibinfo {author}
{\bibfnamefont {L.}~\bibnamefont {Ludhova}}, \bibinfo {author} {\bibfnamefont
{S.}~\bibnamefont {Mertens}}, \bibinfo {author} {\bibfnamefont
{A.}~\bibnamefont {Schukraft}},\ and\ \bibinfo {author} {\bibfnamefont
{M.}~\bibnamefont {Wurm}},\ }\href
{https://doi.org/10.1016/j.ppnp.2019.103736} {\bibfield {journal} {\bibinfo
{journal} {Prog. Part. Nucl. Phys.}\ }\textbf {\bibinfo {volume} {111}},\
\bibinfo {pages} {103736} (\bibinfo {year} {2020})},\ \Eprint
{https://arxiv.org/abs/1906.01739} {arXiv:1906.01739 [hep-ex]} \BibitemShut
{NoStop}%
\bibitem [{\citenamefont {Serebrov}\ \emph {et~al.}(2019)\citenamefont
{Serebrov} \emph {et~al.}}]{Serebrov:2018vdw}%
\BibitemOpen
\bibfield {author} {\bibinfo {author} {\bibfnamefont {A.~P.}\ \bibnamefont
{Serebrov}} \emph {et~al.} (\bibinfo {collaboration} {NEUTRINO-4}),\ }\href
{https://doi.org/10.1134/S0021364019040040} {\bibfield {journal} {\bibinfo
{journal} {Pisma Zh. Eksp. Teor. Fiz.}\ }\textbf {\bibinfo {volume} {109}},\
\bibinfo {pages} {209} (\bibinfo {year} {2019})},\ \Eprint
{https://arxiv.org/abs/1809.10561} {arXiv:1809.10561 [hep-ex]} \BibitemShut
{NoStop}%
\bibitem [{\citenamefont {Giunti}\ \emph {et~al.}(2021)\citenamefont {Giunti},
\citenamefont {Li}, \citenamefont {Ternes},\ and\ \citenamefont
{Zhang}}]{Giunti:2021iti}%
\BibitemOpen
\bibfield {author} {\bibinfo {author} {\bibfnamefont {C.}~\bibnamefont
{Giunti}}, \bibinfo {author} {\bibfnamefont {Y.~F.}\ \bibnamefont {Li}},
\bibinfo {author} {\bibfnamefont {C.~A.}\ \bibnamefont {Ternes}},\ and\
\bibinfo {author} {\bibfnamefont {Y.~Y.}\ \bibnamefont {Zhang}},\ }\href
{https://doi.org/10.1016/j.physletb.2021.136214} {\bibfield {journal}
{\bibinfo {journal} {Phys. Lett. B}\ }\textbf {\bibinfo {volume} {816}},\
\bibinfo {pages} {136214} (\bibinfo {year} {2021})},\ \Eprint
{https://arxiv.org/abs/2101.06785} {arXiv:2101.06785 [hep-ph]} \BibitemShut
{NoStop}%
\bibitem [{\citenamefont {Ohlsson}\ \emph {et~al.}(2010)\citenamefont
{Ohlsson}, \citenamefont {Popa},\ and\ \citenamefont
{Zhang}}]{Ohlsson:2010ca}%
\BibitemOpen
\bibfield {author} {\bibinfo {author} {\bibfnamefont {T.}~\bibnamefont
{Ohlsson}}, \bibinfo {author} {\bibfnamefont {C.}~\bibnamefont {Popa}},\ and\
\bibinfo {author} {\bibfnamefont {H.}~\bibnamefont {Zhang}},\ }\href
{https://doi.org/10.1016/j.physletb.2010.07.053} {\bibfield {journal}
{\bibinfo {journal} {Phys. Lett. B}\ }\textbf {\bibinfo {volume} {692}},\
\bibinfo {pages} {257} (\bibinfo {year} {2010})},\ \Eprint
{https://arxiv.org/abs/1007.0106} {arXiv:1007.0106 [hep-ph]} \BibitemShut
{NoStop}%
\bibitem [{\citenamefont {Parke}\ and\ \citenamefont
{Ross-Lonergan}(2016)}]{Parke:2015goa}%
\BibitemOpen
\bibfield {author} {\bibinfo {author} {\bibfnamefont {S.}~\bibnamefont
{Parke}}\ and\ \bibinfo {author} {\bibfnamefont {M.}~\bibnamefont
{Ross-Lonergan}},\ }\href {https://doi.org/10.1103/PhysRevD.93.113009}
{\bibfield {journal} {\bibinfo {journal} {Phys. Rev. D}\ }\textbf {\bibinfo
{volume} {93}},\ \bibinfo {pages} {113009} (\bibinfo {year} {2016})},\
\Eprint {https://arxiv.org/abs/1508.05095} {arXiv:1508.05095 [hep-ph]}
\BibitemShut {NoStop}%
\bibitem [{\citenamefont {de~Gouv\^ea}\ and\ \citenamefont
{Kobach}(2016)}]{deGouvea:2015euy}%
\BibitemOpen
\bibfield {author} {\bibinfo {author} {\bibfnamefont {A.}~\bibnamefont
{de~Gouv\^ea}}\ and\ \bibinfo {author} {\bibfnamefont {A.}~\bibnamefont
{Kobach}},\ }\href {https://doi.org/10.1103/PhysRevD.93.033005} {\bibfield
{journal} {\bibinfo {journal} {Phys. Rev. D}\ }\textbf {\bibinfo {volume}
{93}},\ \bibinfo {pages} {033005} (\bibinfo {year} {2016})},\ \Eprint
{https://arxiv.org/abs/1511.00683} {arXiv:1511.00683 [hep-ph]} \BibitemShut
{NoStop}%
\bibitem [{\citenamefont {Miranda}\ \emph {et~al.}(2016)\citenamefont
{Miranda}, \citenamefont {Tortola},\ and\ \citenamefont
{Valle}}]{Miranda:2016wdr}%
\BibitemOpen
\bibfield {author} {\bibinfo {author} {\bibfnamefont {O.~G.}\ \bibnamefont
{Miranda}}, \bibinfo {author} {\bibfnamefont {M.}~\bibnamefont {Tortola}},\
and\ \bibinfo {author} {\bibfnamefont {J.~W.~F.}\ \bibnamefont {Valle}},\
}\href {https://doi.org/10.1103/PhysRevLett.117.061804} {\bibfield {journal}
{\bibinfo {journal} {Phys. Rev. Lett.}\ }\textbf {\bibinfo {volume} {117}},\
\bibinfo {pages} {061804} (\bibinfo {year} {2016})},\ \Eprint
{https://arxiv.org/abs/1604.05690} {arXiv:1604.05690 [hep-ph]} \BibitemShut
{NoStop}%
\bibitem [{\citenamefont {Fernandez-Martinez}\ \emph
{et~al.}(2016)\citenamefont {Fernandez-Martinez}, \citenamefont
{Hernandez-Garcia},\ and\ \citenamefont
{Lopez-Pavon}}]{Fernandez-Martinez:2016lgt}%
\BibitemOpen
\bibfield {author} {\bibinfo {author} {\bibfnamefont {E.}~\bibnamefont
{Fernandez-Martinez}}, \bibinfo {author} {\bibfnamefont {J.}~\bibnamefont
{Hernandez-Garcia}},\ and\ \bibinfo {author} {\bibfnamefont {J.}~\bibnamefont
{Lopez-Pavon}},\ }\href {https://doi.org/10.1007/JHEP08(2016)033} {\bibfield
{journal} {\bibinfo {journal} {JHEP}\ }\textbf {\bibinfo {volume} {08}},\
\bibinfo {pages} {033}},\ \Eprint {https://arxiv.org/abs/1605.08774}
{arXiv:1605.08774 [hep-ph]} \BibitemShut {NoStop}%
\bibitem [{\citenamefont {P\"as}\ and\ \citenamefont
{Sicking}(2017)}]{Pas:2016qbg}%
\BibitemOpen
\bibfield {author} {\bibinfo {author} {\bibfnamefont {H.}~\bibnamefont
{P\"as}}\ and\ \bibinfo {author} {\bibfnamefont {P.}~\bibnamefont
{Sicking}},\ }\href {https://doi.org/10.1103/PhysRevD.95.075004} {\bibfield
{journal} {\bibinfo {journal} {Phys. Rev. D}\ }\textbf {\bibinfo {volume}
{95}},\ \bibinfo {pages} {075004} (\bibinfo {year} {2017})},\ \Eprint
{https://arxiv.org/abs/1611.08450} {arXiv:1611.08450 [hep-ph]} \BibitemShut
{NoStop}%
\bibitem [{\citenamefont {Martinez-Soler}\ and\ \citenamefont
{Minakata}(2020)}]{Martinez-Soler:2018lcy}%
\BibitemOpen
\bibfield {author} {\bibinfo {author} {\bibfnamefont {I.}~\bibnamefont
{Martinez-Soler}}\ and\ \bibinfo {author} {\bibfnamefont {H.}~\bibnamefont
{Minakata}},\ }\href {https://doi.org/10.1093/ptep/ptaa062} {\bibfield
{journal} {\bibinfo {journal} {PTEP}\ }\textbf {\bibinfo {volume} {2020}},\
\bibinfo {pages} {063B01} (\bibinfo {year} {2020})},\ \Eprint
{https://arxiv.org/abs/1806.10152} {arXiv:1806.10152 [hep-ph]} \BibitemShut
{NoStop}%
\bibitem [{\citenamefont {Coutinho}\ \emph {et~al.}(2020)\citenamefont
{Coutinho}, \citenamefont {Crivellin},\ and\ \citenamefont
{Manzari}}]{Coutinho:2019aiy}%
\BibitemOpen
\bibfield {author} {\bibinfo {author} {\bibfnamefont {A.~M.}\ \bibnamefont
{Coutinho}}, \bibinfo {author} {\bibfnamefont {A.}~\bibnamefont
{Crivellin}},\ and\ \bibinfo {author} {\bibfnamefont {C.~A.}\ \bibnamefont
{Manzari}},\ }\href {https://doi.org/10.1103/PhysRevLett.125.071802}
{\bibfield {journal} {\bibinfo {journal} {Phys. Rev. Lett.}\ }\textbf
{\bibinfo {volume} {125}},\ \bibinfo {pages} {071802} (\bibinfo {year}
{2020})},\ \Eprint {https://arxiv.org/abs/1912.08823} {arXiv:1912.08823
[hep-ph]} \BibitemShut {NoStop}%
\bibitem [{\citenamefont {Ellis}\ \emph {et~al.}(2020)\citenamefont {Ellis},
\citenamefont {Kelly},\ and\ \citenamefont {Li}}]{Ellis:2020hus}%
\BibitemOpen
\bibfield {author} {\bibinfo {author} {\bibfnamefont {S.~A.~R.}\
\bibnamefont {Ellis}}, \bibinfo {author} {\bibfnamefont {K.~J.}\ \bibnamefont
{Kelly}},\ and\ \bibinfo {author} {\bibfnamefont {S.~W.}\ \bibnamefont
{Li}},\ }\href {https://doi.org/10.1007/JHEP12(2020)068} {\bibfield
{journal} {\bibinfo {journal} {JHEP}\ }\textbf {\bibinfo {volume} {12}},\
\bibinfo {pages} {068}},\ \Eprint {https://arxiv.org/abs/2008.01088}
{arXiv:2008.01088 [hep-ph]} \BibitemShut {NoStop}%
\bibitem [{\citenamefont {Chakraborty}\ \emph {et~al.}(2021)\citenamefont
{Chakraborty}, \citenamefont {Goswami},\ and\ \citenamefont
{Long}}]{Chakraborty:2020brc}%
\BibitemOpen
\bibfield {author} {\bibinfo {author} {\bibfnamefont {K.}~\bibnamefont
{Chakraborty}}, \bibinfo {author} {\bibfnamefont {S.}~\bibnamefont
{Goswami}},\ and\ \bibinfo {author} {\bibfnamefont {K.}~\bibnamefont
{Long}},\ }\href {https://doi.org/10.1103/PhysRevD.103.075009} {\bibfield
{journal} {\bibinfo {journal} {Phys. Rev. D}\ }\textbf {\bibinfo {volume}
{103}},\ \bibinfo {pages} {075009} (\bibinfo {year} {2021})},\ \Eprint
{https://arxiv.org/abs/2007.03321} {arXiv:2007.03321 [hep-ph]} \BibitemShut
{NoStop}%
\bibitem [{\citenamefont {Hu}\ \emph {et~al.}(2021)\citenamefont {Hu},
\citenamefont {Ling}, \citenamefont {Tang},\ and\ \citenamefont
{Wang}}]{Hu:2020oba}%
\BibitemOpen
\bibfield {author} {\bibinfo {author} {\bibfnamefont {Z.}~\bibnamefont
{Hu}}, \bibinfo {author} {\bibfnamefont {J.}~\bibnamefont {Ling}}, \bibinfo
{author} {\bibfnamefont {J.}~\bibnamefont {Tang}},\ and\ \bibinfo {author}
{\bibfnamefont {T.}~\bibnamefont {Wang}},\ }\href
{https://doi.org/10.1007/JHEP01(2021)124} {\bibfield {journal} {\bibinfo
{journal} {JHEP}\ }\textbf {\bibinfo {volume} {01}},\ \bibinfo {pages}
{124}},\ \Eprint {https://arxiv.org/abs/2008.09730} {arXiv:2008.09730
[hep-ph]} \BibitemShut {NoStop}%
\bibitem [{\citenamefont {Meloni}\ \emph {et~al.}(2010)\citenamefont {Meloni},
\citenamefont {Ohlsson}, \citenamefont {Winter},\ and\ \citenamefont
{Zhang}}]{Meloni:2009cg}%
\BibitemOpen
\bibfield {author} {\bibinfo {author} {\bibfnamefont {D.}~\bibnamefont
{Meloni}}, \bibinfo {author} {\bibfnamefont {T.}~\bibnamefont {Ohlsson}},
\bibinfo {author} {\bibfnamefont {W.}~\bibnamefont {Winter}},\ and\ \bibinfo
{author} {\bibfnamefont {H.}~\bibnamefont {Zhang}},\ }\href
{https://doi.org/10.1007/JHEP04(2010)041} {\bibfield {journal} {\bibinfo
{journal} {JHEP}\ }\textbf {\bibinfo {volume} {04}},\ \bibinfo {pages}
{041}},\ \Eprint {https://arxiv.org/abs/0912.2735} {arXiv:0912.2735 [hep-ph]}
\BibitemShut {NoStop}%
\bibitem [{\citenamefont {Dutta}\ \emph {et~al.}(2017)\citenamefont {Dutta},
\citenamefont {Ghoshal},\ and\ \citenamefont {Roy}}]{Dutta:2016czj}%
\BibitemOpen
\bibfield {author} {\bibinfo {author} {\bibfnamefont {D.}~\bibnamefont
{Dutta}}, \bibinfo {author} {\bibfnamefont {P.}~\bibnamefont {Ghoshal}},\
and\ \bibinfo {author} {\bibfnamefont {S.}~\bibnamefont {Roy}},\ }\href
{https://doi.org/10.1016/j.nuclphysb.2017.04.018} {\bibfield {journal}
{\bibinfo {journal} {Nucl. Phys. B}\ }\textbf {\bibinfo {volume} {920}},\
\bibinfo {pages} {385} (\bibinfo {year} {2017})},\ \Eprint
{https://arxiv.org/abs/1609.07094} {arXiv:1609.07094 [hep-ph]} \BibitemShut
{NoStop}%
\bibitem [{\citenamefont {Li}\ \emph {et~al.}(2018)\citenamefont {Li},
\citenamefont {Xing},\ and\ \citenamefont {Zhu}}]{Li:2018jgd}%
\BibitemOpen
\bibfield {author} {\bibinfo {author} {\bibfnamefont {Y.-F.}\ \bibnamefont
{Li}}, \bibinfo {author} {\bibfnamefont {Z.-z.}\ \bibnamefont {Xing}},\ and\
\bibinfo {author} {\bibfnamefont {J.-y.}\ \bibnamefont {Zhu}},\ }\href
{https://doi.org/10.1016/j.physletb.2018.05.079} {\bibfield {journal}
{\bibinfo {journal} {Phys. Lett. B}\ }\textbf {\bibinfo {volume} {782}},\
\bibinfo {pages} {578} (\bibinfo {year} {2018})},\ \Eprint
{https://arxiv.org/abs/1802.04964} {arXiv:1802.04964 [hep-ph]} \BibitemShut
{NoStop}%
\bibitem [{\citenamefont {Soumya}\ and\ \citenamefont
{Rukmani}(2018)}]{Soumya:2018nkw}%
\BibitemOpen
\bibfield {author} {\bibinfo {author} {\bibfnamefont {C.}~\bibnamefont
{Soumya}}\ and\ \bibinfo {author} {\bibfnamefont {M.}~\bibnamefont
{Rukmani}},\ }\href {https://doi.org/10.1088/1361-6471/aad2cc} {\bibfield
{journal} {\bibinfo {journal} {J. Phys. G}\ }\textbf {\bibinfo {volume}
{45}},\ \bibinfo {pages} {095003} (\bibinfo {year} {2018})}\BibitemShut
{NoStop}%
\bibitem [{\citenamefont {Miranda}\ \emph {et~al.}(2021)\citenamefont
{Miranda}, \citenamefont {Pasquini}, \citenamefont {Rahaman},\ and\
\citenamefont {Razzaque}}]{Miranda:2019ynh}%
\BibitemOpen
\bibfield {author} {\bibinfo {author} {\bibfnamefont {L.~S.}\ \bibnamefont
{Miranda}}, \bibinfo {author} {\bibfnamefont {P.}~\bibnamefont {Pasquini}},
\bibinfo {author} {\bibfnamefont {U.}~\bibnamefont {Rahaman}},\ and\ \bibinfo
{author} {\bibfnamefont {S.}~\bibnamefont {Razzaque}},\ }\href
{https://doi.org/10.1140/epjc/s10052-021-09227-0} {\bibfield {journal}
{\bibinfo {journal} {Eur. Phys. J. C}\ }\textbf {\bibinfo {volume} {81}},\
\bibinfo {pages} {444} (\bibinfo {year} {2021})},\ \Eprint
{https://arxiv.org/abs/1911.09398} {arXiv:1911.09398 [hep-ph]} \BibitemShut
{NoStop}%
\bibitem [{\citenamefont {Kelly}\ \emph {et~al.}(2021)\citenamefont {Kelly},
\citenamefont {Machado}, \citenamefont {Parke}, \citenamefont
{Perez~Gonzalez},\ and\ \citenamefont {Zukanovich-Funchal}}]{Kelly:2020fkv}%
\BibitemOpen
\bibfield {author} {\bibinfo {author} {\bibfnamefont {K.~J.}\ \bibnamefont
{Kelly}}, \bibinfo {author} {\bibfnamefont {P.~A.}\ \bibnamefont {Machado}},
\bibinfo {author} {\bibfnamefont {S.~J.}\ \bibnamefont {Parke}}, \bibinfo
{author} {\bibfnamefont {Y.~F.}\ \bibnamefont {Perez~Gonzalez}},\ and\
\bibinfo {author} {\bibfnamefont {R.}~\bibnamefont {Zukanovich-Funchal}},\
}\href {https://doi.org/10.1103/PhysRevD.103.013004} {\bibfield {journal}
{\bibinfo {journal} {Phys. Rev. D}\ }\textbf {\bibinfo {volume} {103}},\
\bibinfo {pages} {013004} (\bibinfo {year} {2021})},\ \Eprint
{https://arxiv.org/abs/2007.08526} {arXiv:2007.08526 [hep-ph]} \BibitemShut
{NoStop}%
\bibitem [{\citenamefont {Astier}\ \emph {et~al.}(2003)\citenamefont {Astier}
\emph {et~al.}}]{Astier:2003gs}%
\BibitemOpen
\bibfield {author} {\bibinfo {author} {\bibfnamefont {P.}~\bibnamefont
{Astier}} \emph {et~al.} (\bibinfo {collaboration} {NOMAD}),\ }\href
{https://doi.org/10.1016/j.physletb.2003.07.029} {\bibfield {journal}
{\bibinfo {journal} {Phys. Lett. B}\ }\textbf {\bibinfo {volume} {570}},\
\bibinfo {pages} {19} (\bibinfo {year} {2003})},\ \Eprint
{https://arxiv.org/abs/hep-ex/0306037} {arXiv:hep-ex/0306037} \BibitemShut
{NoStop}%
\bibitem [{\citenamefont {Avvakumov}\ \emph {et~al.}(2002)\citenamefont
{Avvakumov} \emph {et~al.}}]{Avvakumov:2002jj}%
\BibitemOpen
\bibfield {author} {\bibinfo {author} {\bibfnamefont {S.}~\bibnamefont
{Avvakumov}} \emph {et~al.} (\bibinfo {collaboration} {NuTeV}),\ }\href
{https://doi.org/10.1103/PhysRevLett.89.011804} {\bibfield {journal}
{\bibinfo {journal} {Phys. Rev. Lett.}\ }\textbf {\bibinfo {volume} {89}},\
\bibinfo {pages} {011804} (\bibinfo {year} {2002})},\ \Eprint
{https://arxiv.org/abs/hep-ex/0203018} {arXiv:hep-ex/0203018} \BibitemShut
{NoStop}%
\bibitem [{\citenamefont {Astier}\ \emph {et~al.}(2001)\citenamefont {Astier}
\emph {et~al.}}]{NOMAD:2001xxt}%
\BibitemOpen
\bibfield {author} {\bibinfo {author} {\bibfnamefont {P.}~\bibnamefont
{Astier}} \emph {et~al.} (\bibinfo {collaboration} {NOMAD}),\ }\href
{https://doi.org/10.1016/S0550-3213(01)00339-X} {\bibfield {journal}
{\bibinfo {journal} {Nucl. Phys. B}\ }\textbf {\bibinfo {volume} {611}},\
\bibinfo {pages} {3} (\bibinfo {year} {2001})},\ \Eprint
{https://arxiv.org/abs/hep-ex/0106102} {arXiv:hep-ex/0106102} \BibitemShut
{NoStop}%
\bibitem [{\citenamefont {Eskut}\ \emph {et~al.}(2008)\citenamefont {Eskut}
\emph {et~al.}}]{CHORUS:2007wlo}%
\BibitemOpen
\bibfield {author} {\bibinfo {author} {\bibfnamefont {E.}~\bibnamefont
{Eskut}} \emph {et~al.} (\bibinfo {collaboration} {CHORUS}),\ }\href
{https://doi.org/10.1016/j.nuclphysb.2007.10.023} {\bibfield {journal}
{\bibinfo {journal} {Nucl. Phys. B}\ }\textbf {\bibinfo {volume} {793}},\
\bibinfo {pages} {326} (\bibinfo {year} {2008})},\ \Eprint
{https://arxiv.org/abs/0710.3361} {arXiv:0710.3361 [hep-ex]} \BibitemShut
{NoStop}%
\bibitem [{\citenamefont {Adamson}\ \emph {et~al.}(2019)\citenamefont {Adamson}
\emph {et~al.}}]{Adamson:2017uda}%
\BibitemOpen
\bibfield {author} {\bibinfo {author} {\bibfnamefont {P.}~\bibnamefont
{Adamson}} \emph {et~al.} (\bibinfo {collaboration} {MINOS+}),\ }\href
{https://doi.org/10.1103/PhysRevLett.122.091803} {\bibfield {journal}
{\bibinfo {journal} {Phys. Rev. Lett.}\ }\textbf {\bibinfo {volume} {122}},\
\bibinfo {pages} {091803} (\bibinfo {year} {2019})},\ \Eprint
{https://arxiv.org/abs/1710.06488} {arXiv:1710.06488 [hep-ex]} \BibitemShut
{NoStop}%
\bibitem [{\citenamefont {Abe}\ \emph {et~al.}(2021)\citenamefont {Abe} \emph
{et~al.}}]{Abe:2021gky}%
\BibitemOpen
\bibfield {author} {\bibinfo {author} {\bibfnamefont {K.}~\bibnamefont
{Abe}} \emph {et~al.} (\bibinfo {collaboration} {T2K}),\ }\href
{https://doi.org/10.1103/PhysRevD.103.112008} {\bibfield {journal} {\bibinfo
{journal} {Phys. Rev. D}\ }\textbf {\bibinfo {volume} {103}},\ \bibinfo
{pages} {112008} (\bibinfo {year} {2021})},\ \Eprint
{https://arxiv.org/abs/2101.03779} {arXiv:2101.03779 [hep-ex]} \BibitemShut
{NoStop}%
\bibitem [{\citenamefont {{Alex Himmel}}(2020)}]{alex_himmel_2020_3959581}%
\BibitemOpen
\bibfield {author} {\bibinfo {author} {\bibnamefont {{Alex Himmel}}},\
}\href {https://doi.org/10.5281/zenodo.3959581} {\bibinfo {title} {{New
Oscillation Results from the {NOvA} Experiment}}} (\bibinfo {year}
{2020})\BibitemShut {NoStop}%
\bibitem [{\citenamefont {{Patrick Dunne}}(2020)}]{patrick_dunne_2020_3959558}%
\BibitemOpen
\bibfield {author} {\bibinfo {author} {\bibnamefont {{Patrick Dunne}}},\
}\href {https://doi.org/10.5281/zenodo.3959558} {\bibinfo {title} {{Latest
Neutrino Oscillation Results from {T2K}}}} (\bibinfo {year}
{2020})\BibitemShut {NoStop}%
\bibitem [{\citenamefont {Acero}\ \emph {et~al.}(2018)\citenamefont {Acero}
\emph {et~al.}}]{NOvA:2018gge}%
\BibitemOpen
\bibfield {author} {\bibinfo {author} {\bibfnamefont {M.}~\bibnamefont
{Acero}} \emph {et~al.} (\bibinfo {collaboration} {NOvA}),\ }\href
{https://doi.org/10.1103/PhysRevD.98.032012} {\bibfield {journal} {\bibinfo
{journal} {Phys.Rev. D}\ }\textbf {\bibinfo {volume} {98}},\ \bibinfo {pages}
{032012} (\bibinfo {year} {2018})},\ \Eprint
{https://arxiv.org/abs/1806.00096} {arXiv:1806.00096 [hep-ex]} \BibitemShut
{NoStop}%
\bibitem [{\citenamefont {Huber}\ \emph {et~al.}(2005)\citenamefont {Huber},
\citenamefont {Lindner},\ and\ \citenamefont {Winter}}]{Huber:2004ka}%
\BibitemOpen
\bibfield {author} {\bibinfo {author} {\bibfnamefont {P.}~\bibnamefont
{Huber}}, \bibinfo {author} {\bibfnamefont {M.}~\bibnamefont {Lindner}},\
and\ \bibinfo {author} {\bibfnamefont {W.}~\bibnamefont {Winter}},\ }\href
{https://doi.org/10.1016/j.cpc.2005.01.003} {\bibfield {journal} {\bibinfo
{journal} {Comput. Phys. Commun.}\ }\textbf {\bibinfo {volume} {167}},\
\bibinfo {pages} {195} (\bibinfo {year} {2005})},\ \Eprint
{https://arxiv.org/abs/hep-ph/0407333} {arXiv:hep-ph/0407333 [hep-ph]}
\BibitemShut {NoStop}%
\bibitem [{\citenamefont {Huber}\ \emph {et~al.}(2007)\citenamefont {Huber},
\citenamefont {Kopp}, \citenamefont {Lindner}, \citenamefont {Rolinec},\ and\
\citenamefont {Winter}}]{Huber:2007ji}%
\BibitemOpen
\bibfield {author} {\bibinfo {author} {\bibfnamefont {P.}~\bibnamefont
{Huber}}, \bibinfo {author} {\bibfnamefont {J.}~\bibnamefont {Kopp}},
\bibinfo {author} {\bibfnamefont {M.}~\bibnamefont {Lindner}}, \bibinfo
{author} {\bibfnamefont {M.}~\bibnamefont {Rolinec}},\ and\ \bibinfo {author}
{\bibfnamefont {W.}~\bibnamefont {Winter}},\ }\href
{https://doi.org/10.1016/j.cpc.2007.05.004} {\bibfield {journal} {\bibinfo
{journal} {Comput. Phys. Commun.}\ }\textbf {\bibinfo {volume} {177}},\
\bibinfo {pages} {432} (\bibinfo {year} {2007})},\ \Eprint
{https://arxiv.org/abs/hep-ph/0701187} {arXiv:hep-ph/0701187 [hep-ph]}
\BibitemShut {NoStop}%
\bibitem [{\citenamefont {Aliaga}\ \emph {et~al.}(2016)\citenamefont {Aliaga}
\emph {et~al.}}]{Aliaga:2016oaz}%
\BibitemOpen
\bibfield {author} {\bibinfo {author} {\bibfnamefont {L.}~\bibnamefont
{Aliaga}} \emph {et~al.} (\bibinfo {collaboration} {MINERvA}),\ }\href
{https://doi.org/10.1103/PhysRevD.94.092005} {\bibfield {journal} {\bibinfo
{journal} {Phys. Rev. D}\ }\textbf {\bibinfo {volume} {94}},\ \bibinfo
{pages} {092005} (\bibinfo {year} {2016})},\ \bibinfo {note} {[Addendum:
Phys.Rev.D 95, 039903 (2017)]},\ \Eprint {https://arxiv.org/abs/1607.00704}
{arXiv:1607.00704 [hep-ex]} \BibitemShut {NoStop}%
\bibitem [{\citenamefont {Barenboim}\ \emph {et~al.}(2020)\citenamefont
{Barenboim}, \citenamefont {Ternes},\ and\ \citenamefont
{T\'ortola}}]{Tortola:2020ncu}%
\BibitemOpen
\bibfield {author} {\bibinfo {author} {\bibfnamefont {G.}~\bibnamefont
{Barenboim}}, \bibinfo {author} {\bibfnamefont {C.~A.}\ \bibnamefont
{Ternes}},\ and\ \bibinfo {author} {\bibfnamefont {M.~A.}\ \bibnamefont
{T\'ortola}},\ }\href {https://doi.org/10.1007/JHEP07(2020)155} {\bibfield
{journal} {\bibinfo {journal} {JHEP}\ }\textbf {\bibinfo {volume} {07}},\
\bibinfo {pages} {155}},\ \Eprint {https://arxiv.org/abs/2005.05975}
{arXiv:2005.05975 [hep-ph]} \BibitemShut {NoStop}%
\bibitem [{\citenamefont {Chatterjee}\ and\ \citenamefont
{Palazzo}(2021)}]{Chatterjee:2020kkm}%
\BibitemOpen
\bibfield {author} {\bibinfo {author} {\bibfnamefont {S.~S.}\ \bibnamefont
{Chatterjee}}\ and\ \bibinfo {author} {\bibfnamefont {A.}~\bibnamefont
{Palazzo}},\ }\href {https://doi.org/10.1103/PhysRevLett.126.051802}
{\bibfield {journal} {\bibinfo {journal} {Phys. Rev. Lett.}\ }\textbf
{\bibinfo {volume} {126}},\ \bibinfo {pages} {051802} (\bibinfo {year}
{2021})},\ \Eprint {https://arxiv.org/abs/2008.04161} {arXiv:2008.04161
[hep-ph]} \BibitemShut {NoStop}%
\bibitem [{\citenamefont {Denton}\ \emph {et~al.}(2021)\citenamefont {Denton},
\citenamefont {Gehrlein},\ and\ \citenamefont {Pestes}}]{Denton:2020uda}%
\BibitemOpen
\bibfield {author} {\bibinfo {author} {\bibfnamefont {P.~B.}\ \bibnamefont
{Denton}}, \bibinfo {author} {\bibfnamefont {J.}~\bibnamefont {Gehrlein}},\
and\ \bibinfo {author} {\bibfnamefont {R.}~\bibnamefont {Pestes}},\ }\href
{https://doi.org/10.1103/PhysRevLett.126.051801} {\bibfield {journal}
{\bibinfo {journal} {Phys. Rev. Lett.}\ }\textbf {\bibinfo {volume} {126}},\
\bibinfo {pages} {051801} (\bibinfo {year} {2021})},\ \Eprint
{https://arxiv.org/abs/2008.01110} {arXiv:2008.01110 [hep-ph]} \BibitemShut
{NoStop}%
\end{thebibliography}

\end{document}